\documentclass[12pt]{article}
\usepackage{graphicx,amssymb,epsfig,epsf} 
 \usepackage{ulem}
\usepackage[usenames]{color}
\usepackage{rotating}
\usepackage{epsfig}
\usepackage{slashed}
\usepackage{amsmath}
\usepackage{pstricks}
\usepackage{color}

\begin{document}

\begin{flushright}
   { OSU-HEP-13-03,\, RECAPP-HRI-2013-008}\\
\end{flushright}

\vskip 2pt

\begin{center}
{\large \bf Collider signatures of mirror fermions in the framework of Left Right Mirror Model}\\
\vskip 20pt
{Shreyashi Chakdar$^{a}$\footnote{chakdar@okstate.edu}, {Kirtiman Ghosh$^{a}$\footnote{kirti.gh@gmail.com}}}, S. Nandi$^{a}$\footnote{s.nandi@okstate.edu} and Santosh Kumar Rai$^{b}$\footnote{skrai@hri.res.in}  \\
\vskip 10pt
{$^a$Department of Physics and Oklahoma Center for High Energy Physics,\\
Oklahoma State University, Stillwater, OK 74078-3072, USA.}\\
\vskip 10pt
{ $^{b}$Regional Centre for Accelerator-based Particle Physics,\\
Harish-Chandra Research Institute,  \\ Chhatnag Road, Jhusi, 
Allahabad 211019, India.
}%Lines break automatically or can be forced with \\

\end{center}

\vskip 5pt
%------------------------------------------------------------------------------
%\centerline{\large\bf Abstract}
%\vskip 15pt
%\noindent
\abstract
{
The idea of left-right symmetry with mirror fermions is very appealing from the symmetry point of view. In this picture, unlike the Standard Model, the symmetry is not only left-right symmetric, but each left handed fermion multiplet is accompanied by new right handed fermion multiplet of opposite chirality. In this work, we consider a gauge symmetry, $SU(3)_c \otimes SU(2)_L\otimes SU(2)_R \otimes U(1)_{Y^\prime}$ supplemented by a discrete $Z_2$ symmetry. Instead of having right handed multiplets for each left handed multiplets of the same fermions as in the usual left-right model, the mirror model include right handed doublets involving new fermions (called mirrors), and similarly for  each right handed  singlet, there are corresponding mirror singlets. Thus the gauge anomaly is naturally absent in this model, and the model also provide a solution for the strong CP problem because of parity conservation. The first stage of symmetry breaking is achieved by a doublet mirror Higgs with a vacuum expectation value $\simeq 10^7$ GeV, needed to explain the neutrino mass $\simeq 10^{-11}$ GeV. The mirror fermions can mix with the ordinary fermions via a scalar which is singlet under the gauge symmetry. In this model, only light mirror particles, having masses in the few hundred GeV range are $\hat{e} , \hat{u}, \hat{d}$ with well-defined spectrum.
$\hat{u}$ and $\hat{d}$ can be pair produced at the LHC, and can be detected as ($u~Z$) and ($d~Z$) resonances. We discuss the signals of these mirror fermions at the LHC, and find that the reach at the LHC can be as large as $m_{\hat q}\simeq 800$ GeV.

}

\section{Introduction}

The non-conservation of parity $P$ (the left-right asymmetry of elementary particles) is well incorporated in the Standard Model (SM) of particle physics. However, it has been considered as an unpleasant feature of the model. One possible way to understand the left-right asymmetry of elementary particles is to enlarge the SM into a left-right (LR) symmetric structure and then, by some spontaneously breaking mechanism, to recover the SM symmetry structure. For instance, in  {left-right} symmetric models \cite{LRmodels}, $SU(2)_R$ interactions are introduced to maintain parity invariance at high energy scales. The symmetry group $SU(2)_L\otimes SU(2)_R \otimes U(1)_{B-L}$ of LR symmetric models can be a part of a grand unified symmetry group such as 
$SO(10)$ \cite{Fritzsch:1974nn} or $E_6$ \cite{GUT} or  superstring inspired models \cite{string}. In the framework of LR symmetric SM, the SM left-handed fermions are placed in the $SU(2)_L$ doublets as they are in the SM while the SM right-handed fermions (with the addition of right-handed neutrinos for the case of leptons) are placed in the $SU(2)_R$ doublets. Subsequently, the LR symmetry is spontaneously broken down to the SM electroweak symmetry using suitable Higgs representations. 

Another interesting solution to the non-conservation of parity in the SM was proposed in a classic paper \cite{LY} by Lee and Yang. They postulated the existence of additional (mirror) fermions of opposite chirality to the SM ones to make the world left-right symmetric at high energies. The advantages of models with mirror fermions to solve some problems in particle physics have already been discussed in the literature. For instance, the existence of mirror neutrinos can naturally explain the smallness of neutrino mass via a see-saw like mechanism \cite{neutrino,Foot:1995pa,Berezhiani:1995yi}. Moreover, it can also be useful for the Dark Matter problem \cite{Berezhiani:1995yi}, neutrino oscillations as well as different neutrino physics anomalies such as solar neutrino deficit and atmospheric neutrino anomaly \cite{Foot:1995pa}. On the other hand, mirror fermions can provide a solution to the strong CP problem if the parity symmetry is imposed \cite{strongCP}. Finally, the existence of mirror particles appear naturally in many extensions of the SM, like GUT and string theories \cite{string_GUT}. The masses of these mirror particles, though unknown, are not experimentally excluded to be at or below the TeV scale. Therefore, it is important to study the phenomenological consequences of the mirror particles in the context of collider experiments, in particular at the  {Large Hadron Collider} (LHC). In this paper, we have investigated the phenomenology of mirror particles in the context of a particular variant of LR symmetric mirror model (LRMM), their associated final state signals, and the discovery potential at the LHC.

In the LRMM we propose in this work, the SM gauge group ($G_{SM}=SU(3)_C\otimes SU(2)_L \otimes U(1)_Y$) is extended to $G_{LR}=SU(3)_C\otimes SU(2)_L\otimes SU(2)_R \otimes U(1)_{Y^{\prime}}$ together with a discrete $Z_2$ symmetry. The SM particle spectrum is also extended to include mirror particles and a real scalar Higgs singlet under both $SU(2)_L$ and $SU(2)_R$. For the fermion sector, the right-handed (left-handed) components of mirror fermions transform as doublets (singlets) under $SU(2)_R$. The SM fermions are singlets under $SU(2)_R$, whereas  doublets  under $SU(2)_L$. Similarly there are mirror singlet fermions corresponding to the SM singlet fermions. Since the fermion representations are exactly mirror symmetric, all triangle anomalies are exactly cancelled with respect to the entire gauge symmetry, the the model is anomaly free. Because of even number of doublets, there is also no gravitational anomaly.
The SM charged fermions are even under the $Z_2$ symmetry, whereas,  {the} corresponding mirror fermions are odd. Therefore, any mass mixing  between SM charged fermions and with mirror partners are forbidden by the $Z_2$ symmetry. The spontaneous symmetry breaking (SSB), $G_{LR} \to G_{SM}$ is realized by introducing a mirror Higgs doublet which is singlet under $SU(2)_L$ and doublet under $SU(2)_R$. Subsequently the SSB, $G_{SM} \to SU(3)_C \otimes U(1)_{EM}$ is achieved via the SM Higgs doublet which is doublet under $SU(2)_L$ and singlet under $SU(2)_R$.
After the SSB, the gauge boson sector of LRMM contains the usual SM gauge bosons (gluon, $W^\pm$ bosons, $Z$ boson and photon) along with the mirror partners of $W^\pm$ and $Z$-boson. The non zero vacuum expectation value (VEV) for a singlet scalar breaks the $Z_2$ symmetry and gives rise to mixing between the SM and mirror fermions.

The parity symmetry in LRMM determines the ratio among the charged mirror fermion masses from the SM charged fermion mass spectrum. In particular, the ratio of the SM fermion mass and the corresponding mirror fermion mass is given by $ {\cal O}(1)\frac{v}{\hat v}$, where $ {\cal O}(1)$ is an order one number, $v\sim 250$ GeV and $\hat v$ are the VEV's for the SM Higgs and mirror Higgs respectively. Connecting the model for generating tiny neutrino masses $\simeq 10^{-11}$ GeV gives $\hat v \sim 10^{7}$ GeV. This gives  TeV scale masses, or few hundred GeV masses for the mirror partners of electron, up and down quarks, namely ${\hat e},~{\hat u}~{\rm and}~{\hat d}$. This makes the model testable at the ongoing LHC and proposed linear electron-positron collider experiments. 
%Apart from the collider phenomenological motivation, $\hat v \sim 10^{7}$ GeV is also motivated from the tiny neutrino mass generation which we have %discussed in the text.

One of the major goals of the LHC experiment is to find new physics beyond the SM. The LHC is a proton-proton collider and thus, the collision processes are overwhelmed by the QCD interactions. Therefore, in the framework of LRMM, the new TeV scale colored particles, namely ${\hat u}~{\rm and}~{\hat d}$ quarks will be copiously pair produced at the LHC. After being produced, ${\hat u}~{\rm and}~{\hat d}$ quarks will decay to the SM particles giving rise to interesting signatures at the LHC. The TeV scale mirror quarks   {are found to} decay into a $Z/W$ boson or a Higgs boson in association with a SM quark.  {This leads to new fermionic resonances as well as  new physics signals in  
two SM gauge bosons + two jet final states. Note that the gauge bosons could be either $Z$ or $W$ and 
the highlight of the signal would be the presence of a clear resonance in the jet+$Z$ and jet+$W$ 
invariant mass distributions. Such a resonance will stand out against any SM background in these final 
states. In this paper we have therefore studied in detail the signal coming from the pair production of the mirror quarks, ${\hat u}~{\rm and}~{\hat d}$ and their subsequent decays in our\begin{small}
•
\end{small} LRMM and compared it 
with the dominant SM background processes.} 

% Thus we get resonances in the jet + Z and jet +W channels. Such resonances are not present in the SM, and will
% be rather easy to observe above the background with a large $p_T$ cut on the jet, and the Z boson. This will be a 
% clear signal for new physics. In addition to the resonant signal, in this paper,  we have also studied  the final state 
% signals arising from  pair production of ${\hat u}~{\rm and}~{\hat d}$ quarks and their subsequent decays. This  
% gives rise to {\it two SM bosons in association with two light quarks} at the LHC.
% In this paper, we have studied {\it two $Z$-boson plus two light quark jets} and {\it one $Z$-boson and one $W$-boson plus two light quark jets} %signals at the LHC as a signature of LRMM.
 
The paper is organized as follows. In Section \ref{sec:model}, we discuss our model and the formalism. Section \ref{sec:pheno} is devoted for the phenomenological implications of the model. Finally, a summary 
of our work, and the conclusions are given in Section \ref{sec:summary}.

\section{Left-Right symmetric mirror model (LRMM) and the formalism}\label{sec:model}

Our LR symmetric mirror model is based on the gauge symmetry $G_{LR}=SU(3)_C\otimes SU(2)_L\otimes SU(2)_R \otimes U(1)_{Y^{\prime}}$ supplemented by a discrete $Z_2$ symmetry. Left-right symmetry, as in the usual left-right model, provides a natural explanation why the parity is violated at low energy. Inclusion of mirror particles gives an alternate realization of the  {LR} symmetry in the 
fermion sector.
%One of the motivation for proposing L-R model with singlet fermions is to solve strong CP problem. It was shown in Ref.~\cite{strongCP} that the %complete invariance of such a model under parity can guarantee a vanishing strong CP phase from the QCD $\theta$-vacuum and thus, solves strong CP %problem.
% In the present version of LRMM, the SM gauge group ($G_{SM}$) is extended to $G_{LR}=SU(3)_C\otimes SU(2)_L\otimes SU(2)_R \otimes U(1)_{Y^{\prime}}$ %with a discrete $Z_2$ symmetry and the matter content  is also enlarged by including new particles with mirror properties. For instance, this model %includes mirror fermions with opposite chirality relative to the SM fermions.
The fermion representations in our model for leptons and quarks in the first family is given by
%sector under the gauge group $G_{LR}$ are presented in the following:
\begin{eqnarray}
l_{L}^{0}={\begin{pmatrix} \nu^0 \\ e^0 \end{pmatrix}}_L\sim (1,2,1,-1) &,& e^0_R \sim (1,1,1,-2)~~,~~ \nu_R^0 \sim(1,1,1,0);\nonumber\\
\hat {l}_{R}^{0}={\begin{pmatrix} \hat{\nu}^0 \\ \hat{e}^0 \end{pmatrix}}_R\sim (1,1,2,-1) &,& \hat {e}^0_L \sim (1,1,1,-2)~~,~~ \hat {\nu}_L^0 \sim(1,1,1,0);\nonumber\\ 
Q_{L}^{0}={\begin{pmatrix} u^0 \\ d^0 \end{pmatrix}}_L\sim (3,2,1,\frac{1}{3}) &,& u^0_R \sim (3,1,1,\frac{4}{3})~~~,~~ d_R^0 \sim(3,1,1,-\frac{2}{3});\nonumber\\
\hat {Q}_{R}^{0}={\begin{pmatrix} \hat{u}^0 \\ \hat{d}^0 \end{pmatrix}}_R\sim (3,1,2,\frac{1}{3}) &,& \hat {u}^0_L \sim (3,1,1,\frac{4}{3})~~~,~~ \hat {d}_L^0 \sim(1,1,1,-\frac{2}{3});
\label{frep} 
\end{eqnarray}
where the bracketed entries correspond to the transformation properties under the symmetries of the group $G_{LR}$. The superscripts ($^0$) denote gauge eigenstates and the hat symbol ($\hat~$) is associated with the mirror fermions. The charge generator is given by: $Q=T_{3L}+T_{3R}+Y^{\prime}/2$. Under the $Z_2$ symmetry, the charged fermions in the mirror sector (denoted by hat) are odd, whereas, the ordinary fermions (denoted by without hat)  are even. The singlet neutrinos, not present in the SM, are even under $Z_2$. These are needed to generate tiny masses for the light observed neutrinos. The fermion representations for the second and third family are identical to the first family. Note that the $Z_2$ is needed so that we do not have mass mixing of the charged fermions between the ordinary fermions and the mirror fermions. This avoids the ordinary charged fermions from getting masses in the first stage of symmetry breaking  {which happens at a high} scale.

Note that in the traditional  {LR} model, the fermion sector is completely symmetric for the ordinary SM fermions. For example, we have $(u,d)_L$ and $(u,d)_R$, and similarly for every  {fermion family}. Another version, proposed in  \cite{LY} is to introduce new fermions to make it  {LR} symmetric, {\it i.e.}  for every $(u,d)_L$, we have new fermions,  $(\hat{u},\hat{d})_R$. Hence it is the left-right mirror model (LRMM). It is this realization that we pursue here.
It was shown in Ref.~\cite{strongCP} that the complete invariance of such a model under parity can guarantee a vanishing strong CP phase from the QCD $\theta$-vacuum and thus, solves strong CP problem.

\subsection{Symmetry breaking and the scalar sector}
In the framework of LRMM, spontaneous symmetry breaking is achieved via the following steps:
 {
\begin{equation}
%SU(3)_C\otimes SU(2)_L\otimes SU(2)_R \otimes U(1)_{Y^{\prime}} \to SU(3)_C\otimes SU(2)_L\otimes U(1)_{Y} \to SU(3)_C\otimes U(1)_{Q},
SU(2)_L\otimes SU(2)_R \otimes U(1)_{Y^{\prime}} \to SU(2)_L\otimes U(1)_{Y} \to U(1)_{Q},
\end{equation}} 
where, $Y/2=T_{3R}+Y^\prime/2$. In order to realize the above SSB, two Higgs doublets (both even under the $Z_2$ symmetry) are required {\it i.e.}, the SM Higgs doublet ($\Phi$) and its mirror partner ($\hat {\Phi}$). The gauge quantum numbers and  VEV's of these Higgs doublets are summarized below:
\begin{eqnarray}
\Phi \sim (1,2,1,1) &,& \hat {\Phi} \sim (1,1,2,1);\nonumber\\
\left < \Phi \right> = \frac{1}{\sqrt 2}{\begin{pmatrix} 0 \\ v \end{pmatrix}} &,& \hat {\left < {\Phi} \right>} = \frac{1}{\sqrt 2}{\begin{pmatrix} 0 \\ \hat {v} \end{pmatrix}} \label{vev}.
\end{eqnarray}
In addition to these two Higgs doublets, we have introduced a singlet (under both $SU(2)_L$ and $SU(2)_R$) real scalar which is odd under the $Z_2$ symmetry: $\chi\sim (1,1,1,0)$. The VEV of $\chi$: $\left< \chi \right>=v_{\chi}$, breaks the $Z_2$ symmetry spontaneously. This enables us to generate mixing between the SM fermions and the mirror fermions. This  {mixing with the SM fermions allows the mirror fermions to decay to 
lighter SM particles after they are pair produced at colliders  such as the LHC,} giving rise to interesting final state signals.  

In order to generate the above structure of VEV's for $\Phi$ and $\hat{\Phi}$, the the LR symmetry has to be broken, otherwise, we will end up with $v=\hat {v}$. The most general scalar potential that develops this pattern of VEV's is given by,
\begin{eqnarray}
V=&-&\left( \mu^2 \Phi^\dagger \Phi + \hat {\mu^2} \hat {\Phi}^\dagger \hat {\Phi}\right ) + \frac{\lambda}{2}\left[ \left(\Phi^\dagger \Phi \right)^2 + \left(\hat {\Phi}^\dagger \hat {\Phi} \right)^2\right ] + \lambda_1 \left(\Phi^\dagger \Phi \right) \left(\hat {\Phi}^\dagger \hat {\Phi} \right)\nonumber\\
&-&\frac{1}{2}\mu_\chi^2\chi^2+\frac{1}{4}\lambda_\chi \chi^4+ \lambda_{\phi\chi}\chi^2\left(\Phi^\dagger\Phi+\hat \Phi^\dagger \hat \Phi\right)
\end{eqnarray}
It is important to  {note} that in the above potential, the terms with $\mu,~\hat{\mu}$ break the parity symmetry softly, {\it i.e.}, only through the dimension-two mass terms of the scalar potential. Note that after the two stages of symmetry breaking, we are left with three neutral scalars, SM like Higgs, $h$,  Mirror Higgs, $\hat{h}$, and a singlet Higgs $\chi$. We consider a solution of the Higgs potential such that $v<< v_{\chi} << \hat{v}$, and so the mixing among these Higgses are negligible.

\subsection{Gauge bosons masses and mixings}
The gauge bosons masses and mixings are obtained from the kinetic terms of the scalars in the Lagrangian:
\begin{equation}
{\cal L} \supset \left({\cal D}_\mu\Phi\right)^\dagger\left({\cal D}^\mu\Phi\right)+\left(\hat {\cal D}_\mu\hat {\Phi}\right)^\dagger\left(\hat {\cal D}^\mu\hat{\Phi}\right),\label{ktl}
\end{equation}
where, ${\cal D}~{\rm and}~\hat {\cal D}$ are the covariant derivatives associated with the SM and mirror sector respectively.
 {\begin{equation}
%{\cal D}_\mu (\hat {\cal D}_\mu)=\partial_\mu + ig_s\frac{\lambda_a}{2}G^a_\mu + ig \frac{\tau_a}{2}W^a_\mu (\hat {W}^a_\mu)+ig^\prime \frac{Y^\prime}{2} B_\mu,
{\cal D}_\mu (\hat {\cal D}_\mu)=\partial_\mu + ig \frac{\tau_a}{2}W^a_\mu (\hat {W}^a_\mu)+ig^\prime \frac{Y^\prime}{2} B_\mu,
\end{equation}}
where, $\lambda_a$'s and $\tau_a$'s are the Gell Mann and Pauli matrices respectively. The gauge bosons and gauge couplings related to the gauge group 
%$SU(3)_C\otimes SU(2)_L\otimes SU(2)_R \otimes U(1)_{Y^{\prime}}$ 
 {$SU(2)_L\otimes SU(2)_R \otimes U(1)_{Y^{\prime}}$}
are respectively %$G_\mu^a$, 
 {$W^a_\mu,~\hat {W}^a_\mu$, $B_\mu$}  and  {$g,~g,~g^\prime$}.  {Note} that to ensure parity symmetry, we have chosen identical gauge coupling for $SU(2)_L$ and $SU(2)_R$.

Substituting the VEV's of Eq.~\ref{vev} in the kinetic terms for the scalars in Eq.~\ref{ktl}, we obtain the masses and mixings of the seven electroweak gauge bosons of this model. The light gauge bosons are denoted as: $W^\pm$, $Z$ and $\gamma$, which are identified with the SM ones, whereas the mirror gauge bosons are denoted by $\hat {W}^\pm$ and $\hat {Z}$. The mass matrix for the charged
gauge bosons is diagonal, with masses:
\begin{equation}
M_{W^\pm}~=~\frac{1}{2}gv~~~~~,~~~~~M_{\hat W^\pm}~=~\frac{1}{2}g\hat v.
\end{equation}    
The mass matrix for the neutral gauge boson sector is not diagonal and in the basis ($W^3,~\hat W^3,~B$), the neutral gauge boson mass matrix is given by,
\begin{equation}
M=\frac{1}{4}{\begin{pmatrix} g^2v^2 & 0 & -gg^\prime v^2 \\ 0 & g^2\hat v^2 & -gg^\prime \hat v^2 \\ -gg^\prime v^2 &  -gg^\prime \hat v^2 & g^{\prime 2}(v^2+\hat v^2) \end{pmatrix}}.\label{nmm}
\end{equation}
This mass matrix can be diagonalized by means of an orthogonal transformation $R$ which connects the weak eigenstates: ($W^3,~\hat W^3,~B$) to the physical mass eigenstates: ($Z,~\hat Z,~\gamma$);
\begin{equation}
{\begin{pmatrix} W^3 \\ \hat W^3 \\ B \end{pmatrix}}=R {\begin{pmatrix} Z \\\hat Z \\ \gamma \end{pmatrix}}.
\end{equation}
We have obtained the eigenvalues and eigenvectors of the matrix in Eq.~\ref{nmm}. The eigenvalues correspond to the masses of the physical states. One eigenstate ($\gamma$) has zero eigenvalue which is identified with the SM photon and the masses of other eigenstates are given by,
\begin{eqnarray}
M_Z^2&=& \frac{1}{4}v^2g^2\frac{g^2+2g^{\prime 2}}{g^2+g^{\prime 2}}\left[1-\frac{g^{\prime 4}}{\left(g^2+g^{\prime 2}\right)^{ {2}}}\epsilon\right],\nonumber\\
M_{\hat Z}^2&=& \frac{1}{4}\hat v^2\left(g^2+g^{\prime 2}\right)\left[1+\frac{g^{\prime 4}}{\left(g^2+g^{\prime 2}\right)^{ {2}}}\epsilon\right],\label{gbmass}
\end{eqnarray}
where, $\epsilon = v^2/\hat v^2$. Since  {we assume that} $\hat v >> v$, the ${\cal O}(\epsilon^2)$ terms in Eq.~\ref{gbmass}  {can be} neglected. The mixing matrix $R$ in the neutral gauge boson sector can be analytically expressed in terms of two mixing angle: $\theta_W$ and $\hat \theta_W$. The angles are defined in the following:
\begin{equation}
{\rm cos}^2\theta_W=\left(\frac{M_W^2}{M_Z^2}\right)_{\epsilon=0}=\frac{g^2+g^{\prime2}}{g^2+2g^{\prime 2}}~~,~~{\rm cos}^2\hat \theta_W=\left(\frac{M_{\hat W}^2}{M_{\hat Z}^2}\right)_{\epsilon=0}=\frac{g^2}{g^2+g^{\prime 2}}.
\end{equation}
The analytic expression for the mixing matrix upto ${\cal O}(\epsilon)$ is given by,
\begin{equation}
R={\begin{pmatrix}
-{\rm cos}\theta_W & -{\rm cos}\hat \theta_W{\rm sin}^2\hat \theta_W \epsilon & {\rm sin}\theta_W \\
{\rm sin}\theta_W {\rm sin} \hat\theta_W \left[ 1+\frac{{\rm cos}^2\hat \theta_W}{{\rm cos}^2\theta_W} \epsilon\right] & -{\rm cos}\hat \theta_W \left[1-{\rm sin}^4\hat\theta_W \epsilon \right] & {\rm sin}\theta_W \\
{\rm sin}\theta_W {\rm cos} \hat\theta_W \left[ 1-\frac{{\rm sin}^2\hat \theta_W}{{\rm cos}\theta_W} \epsilon\right] & {\rm sin}\hat \theta_W \left[1+{\rm sin}^2\hat\theta_W {\rm cos}^2\hat\theta_W \epsilon \right] & {\rm cos}\theta_W {\rm cos}\hat \theta_W
\end{pmatrix}}
\end{equation}  
It is important to  {note} that in the limit $\epsilon=0$, one recovers the SM gauge boson couplings. The couplings of our theory are related to the electric charge ($e$) by,
\begin{equation}
g=\frac{e}{{\rm sin}\theta_W},~~g^\prime=\frac{e}{{\rm cos}\theta_W {\rm cos}\hat \theta_W},~~
{\rm which~ implies},~~
\frac{1}{e^2}=\frac{2}{g^2}+\frac{1}{g^{\prime 2}}.
\end{equation}

Note that there are only two independent gauge couplings in the theory which we express in terms of 
$e$ and ${\rm cos}\theta_W$  {and therefore} $\hat{\theta}_W$ is not an independent angle, but 
is related to $\theta_W$  {as ${\rm sin} \hat \theta_W={\rm tan} \theta_W$}. 

\subsection{Fermion mass and mixing}
{\bf Charged fermion sector:}\\
The charged fermion mass Lagrangian includes Yukawa terms for the SM fermions and its mirror partners. Mass terms between the singlet SM fermions and mirror fermions are forbidden by the $Z_2$ symmetry. However, the Yukawa interactions between the singlet SM fermions and mirror fermions with the singlet scalar $\chi$ are allowed. The Lagrangian invariant under our gauge symmetry as well as the $Z_2$ symmetry for the down quark and its mirror partner is given by,
\begin{eqnarray}
{\cal L} &\supset& y_d \left(\bar Q^0_L \Phi d^0_R + \bar {\hat Q}^0_R \hat \Phi \hat d^0_L\right) + h_d ~\chi \bar d_R \hat d_L + {\rm h.c.}\nonumber\\
&\supset& {\begin{pmatrix} \bar d_L^0 & \bar {\hat d}^0_L \end{pmatrix}}{\begin{pmatrix}\frac{y_d v}{\sqrt 2} & 0 \\ M^*_{d\hat d}  & \frac{y^*_d \hat v}{\sqrt 2}\end{pmatrix}}{ {\begin{pmatrix} d^0_R \\  {\hat d}^0_R \end{pmatrix}}}~+~{\rm h.c.},
\end{eqnarray}  
where, $y_d$ and $h_d$ are the Yukawa couplings for the SM $d$-quark and $M_{d \hat d}=h_d v_\chi$. To ensure LR symmetry, we have used  {the} same Yukawa coupling for the ordinary and  the mirror sector. Notice that the Yukawa terms involving $\chi$ introduce mixing between SM and mirror fermions. The charged fermion mass matrix can be diagonalized via bi-unitary transformation by introducing two mixing angles. The charged fermion mass (physical) eigenstates are related to the gauge eigenstates by the following relation:
\begin{equation}
{\begin{pmatrix} f^0 \\  {\hat f}^0 \end{pmatrix}}_{L,R}~=~{\begin{pmatrix}{\rm cos}\theta^f & {\rm sin}\theta^f \\-{\rm sin}\theta^f & {\rm cos}\theta^f \end{pmatrix}}_{L,R}{\begin{pmatrix}  f \\  {\hat f} \end{pmatrix}}_{L,R}
\end{equation}
where, $f_{L,R}$ can be identified with the L and R-handed component of the SM fermions and $\hat f_{L,R}$ corresponds to the heavy mirror fermions. The masses and mixing angles are given by:
\begin{eqnarray}
m_{f}~=~\frac{y_f v}{\sqrt 2}&,& m_{\hat f}~=~\sqrt{\frac{y^2_{ {f}}\hat v^2+ {2}M_{f\hat f}^2}{2}};\nonumber\\
{\rm tan}2\theta^f_{R}~=~\frac{2{\sqrt{2}}y_f M_{f\hat f}\hat v}{y_f^2\left(v^2-\hat v^2\right)+ {2}M_{f\hat f}^2}&,& {\rm tan}2\theta^f_{L}~=~\frac{2{\sqrt{2}}y_f M_{f\hat f}v}{y_f^2\left(v^2-\hat v^2\right)- {2}M_{f\hat f}^2}.
\label{tan}
\end{eqnarray}

\medskip
\noindent{\bf Neutrino Sector}:\\
The SM and singlet neutrinos (both in the ordinary and the mirror sector) are even under the $Z_2$ symmetry. Therefore, the mass terms between $SU(2)_L$ and $SU(2)_R$ singlet neutrinos are allowed.
The Lagrangian allowed by our gauge symmetry and  {respecting the} discrete $Z_2$ symmetry is given by
 {\begin{eqnarray*}
{\cal L} &\supset& f_{\nu} \left(\bar l^0_L \Phi \nu^0_R + \bar {\hat l}^0_R \hat \Phi \hat \nu^0_L\right) +M  \nu^{0T}_R C^{-1}  \nu^0_R + M \bar{\nu}^0_R \hat \nu^0_L
+ M \hat \nu^{0T}_L C^{-1} \hat \nu^0_L + {\rm h.c.}
%&\supset& {\begin{pmatrix} \bar d_L^0 & \bar {\hat d}^0_L \end{pmatrix}}{\begin{pmatrix}\frac{y_d v}{\sqrt 2} & 0 \\ M^*_{d\hat d}  & \frac{y^*_d \hat %v}{\sqrt 2}\end{pmatrix}}{\begin{pmatrix} \bar d^0_R \\ \bar {\hat d}^0_R \end{pmatrix}}~+~{\rm h.c.},
\end{eqnarray*}}
%where l's are lepton doublets.  
where $f_{\nu}$ is the neutrino Yukawa coupling, and M is the singlet neutrino mass of order $\hat{v}$.
The neutrino mass matrix with both Dirac  {mass ($m=f_\nu v/\sqrt{2}$ and $m^\prime=f_\nu \hat v/\sqrt{2}$)} and Majorana mass  {($M$)} terms in $(\nu^0_L,\nu^0_R,\hat \nu^0_R, \hat \nu^0_L)$ basis is given by,
\begin{equation}
{\begin{pmatrix}
0 & m & 0 & 0 \\
m & M & 0 & M \\
0 & 0 & 0 & m' \\
0 & M & m'& M\\
\end{pmatrix}}.
\end{equation}  
%where, $f$ is the  neutrino Yukawa coupling and $M$ is the singlet neutrino mass which should be of order $\hat{v}$.  Assuming $M \sim \hat v$, 
The order of magnitude for the eigenvalues of the neutrino mass matrix are given 
\begin{equation}
-m^2/2M,~m^\prime \sqrt{2},~{m^{\prime}}^2/2m,~2M.
\end{equation}
Thus to generate a light neutrino mass  $\simeq 10^{-11}$ GeV with a  {Yukawa coupling strength} of $f_\nu \sim 10^{-4}$  (which is similar to the Yukawa coupling of the electron), we need  $\hat{v}\sim 10^{7}$ GeV. This $\hat{v}\sim 10^{7}$ scale then determines the masses of the mirror fermions. For the first family, the mirror fermion masses then  {come out to be} in the few hundred GeV to TeV range. 
 {Note that to fit the neutrino mass and mixing angles to experimental data would require a more detailed analysis of the neutrino sector which we leave for future studies. Another realization with a mirror like symmetry 
to generate neutrino masses was considered in Ref~\cite{Hung:2006ap}. }

\begin{table}[t!]
\begin{center}
\begin{tabular}{||c|c||c|c||}
\hline \hline
$f$ & $f^\prime$ & $A_{ff^\prime}^{W}$ & $A_{ff^\prime}^{\hat W}$ \\\hline\hline
$d$ & $u$ & ${\rm cos}^2 \theta_L$ & ${\rm sin}^2\theta_R$ \\
$d$ & $\hat u$ & ${\rm cos}\theta_L{\rm sin}\theta_L$ & -${\rm cos}\theta_R {\rm sin}\theta_R$ \\
$\hat d$ & $u$ & ${\rm cos}\theta_L{\rm sin}\theta_L$ & -${\rm cos}\theta_R {\rm sin}\theta_R$ \\
$\hat d$ & $\hat u$ & ${\rm sin}^2 \theta_L$ & ${\rm cos}^2\theta_R$ \\\hline\hline
\end{tabular}
\end{center}
\caption{\small{Analytical expressions for $A_{ff^\prime}^{W}$ and $A_{ff^\prime}^{\hat W}$. Note that we have assumed $V_{ud}=1$. We have also assumed fermion mixing angles ($\theta_L$ and $\theta_R$) are same for up and down flavor.}}
\label{tab:cc}
\end{table}
\section{Phenomenology}\label{sec:pheno}
In this section, we discuss the  {collider} phenomenology of the LRMM. Before going into the details of the collider signatures of LRMM, we  {first need to} study the properties of mirror fermions and bosons. From the point of view of collider phenomenology, we are interested in the interactions between SM particles and mirror particles  {which give the production and decay properties of the mirror particles}. The Lagrangian for the charge currents with $W^\pm$ and $\hat W^\pm$ boson contributions are given by, 
\begin{equation}
{\cal L}_{CC}=-\frac{g}{2\sqrt 2}\bar f \gamma^\mu \left[A_{ff^\prime}^{W}(1-\gamma^5)W_\mu^{-}+A_{ff^\prime}^{\hat{W}}(1+\gamma^5)\hat{W}_\mu^{-}\right]f^\prime,
\label{eq:cc}
\end{equation}
where the coefficients $A_{ff^\prime}^{W}~{\rm and}~A_{ff^\prime}^{\hat{W}}$ depend on the charged fermion mixing angles: $\theta_L~{\rm and}~\theta_R$. The analytical expressions for these coefficients are presented in Table~\ref{tab:cc}\footnote{Fermion mixing angles ($\theta_L~{\rm and}~\theta_R$) depend on the Yukawa coupling of the corresponding fermion. Therefore, the mixing angles are different for up and down flavor. However, we have used the same symbol for the mixing angles of up and down quarks.} for up and down flavored SM and mirror fermions. The neutral current interactions of fermions with neutral gauge bosons ($\gamma,~Z$ and $\hat Z$-bosons) are described by the following Lagrangian. 
\begin{eqnarray}
{\cal L}_{NC} = &-& e Q_f \bar f \gamma^\mu A_\mu f \nonumber\\
&-&\frac{1}{6}\frac{g}{{\rm cos}^3\theta_W}\bar f \gamma^\mu \left[A_{ff^\prime}^Z \frac{1-\gamma^5}{2}+B_{ff^\prime}^Z \frac{1+\gamma^5}{2}\right] Z_\mu f^\prime \nonumber\\
&-&\frac{1}{6}\frac{g}{{\rm cos}^3\theta_W \sqrt{{\rm cos}2\theta_W}}\bar f \gamma^\mu \left[A_{ff^\prime}^{\hat Z} \frac{1-\gamma^5}{2}+B_{ff^\prime}^{\hat Z} \frac{1+\gamma^5}{2}\right] {\hat Z}_\mu f^\prime ,
\label{eq:NC}
\end{eqnarray}
where, $e$ is electron charge and $Q_f$ is the charge of fermion $f$. For up and down flavored SM and mirror fermions, analytical expressions upto $O(\epsilon)$ for the coefficients 
\begin{table}[th!]
\small{
\begin{center}
\begin{tabular}{||c|c||c|c||}
\hline \hline
$f$ & $f^\prime$ & $A_{ff^\prime}^{Z}$ & $B_{ff^\prime}^{Z}$ \\\hline\hline
$d$ & $d$ & $3{\rm cos}^2\theta_L{\rm cos}^2\theta_W-2{\rm cos}^2\theta_W{\rm sin}^2\theta_W$ & $-2 {\rm cos}^2{\theta_W}{\rm sin}^2\theta_W-3 {\rm sin}^2{\theta_R}    {\rm sin}^2{\theta_W}    \sqrt{{\rm cos}2\theta_W}\epsilon$\\
& & $-(1-3{\rm sin}^2\theta_L){\rm sin}^2\theta_W\epsilon$ & $+(2-3 {\rm sin}^2{\theta_R})   {\rm sin}^3{\theta_W}    \epsilon $\\\hline
$d$ & $\hat d$ & $3{\rm cos}^2{\theta_W}    {\rm sin}\theta_L   {\rm cos}{\theta_L}- 3{\rm sin}\theta_L   {\rm sin}^3{\theta_W}    {\rm cos}{\theta_L}   \epsilon$ & $3{\rm sin}{\theta_R}   {\rm sin}^2{\theta_W}   {\rm cos}{\theta_R}   \sqrt{{\rm cos}2\theta_W}\epsilon$\\
& & & $+3{\rm sin}{\theta_R}   {\rm sin}^3{\theta_W}    {\rm cos}{\theta_R}   \epsilon$ \\\hline
$\hat d$ & $\hat d$ & $3 {\rm cos}^2{\theta_W}    {\rm sin}^2\theta_L-2 {\rm cos}^2{\theta_W}    {\rm sin}^2{\theta_W}$ & $-3 {\rm cos}^2{\theta_R}    {\rm sin}^2{\theta_W}    \sqrt{{\rm cos}2\theta_W}\epsilon$ \\
&& $+ (2-3 {\rm sin}^2\theta_L)   {\rm sin}^3{\theta_W}    \epsilon$ & $- (1-3 {\rm sin}^2{\theta_R})   {\rm sin}^3{\theta_W}    \epsilon$ \\\hline
$u$ & $u$ & $-3 {\rm cos}^2{\theta_L}    {\rm cos}^2{\theta_W}+4 {\rm cos}^2{\theta_W}    {\rm sin}^2{\theta_W}$ & $+4 {\rm cos}^2{\theta_W}    {\rm sin}^2{\theta_W}+3 {\rm sin}^2{\theta_R}    {\rm sin}^2{\theta_W}    \sqrt{{\rm cos}2\theta_W}\epsilon$\\
& & $- (1+3 {\rm sin}^2\theta_L)   {\rm sin}^3{\theta_W}    \epsilon$ & $- (4-3 {\rm sin}^2{\theta_R})   {\rm sin}^3{\theta_W}    \epsilon$\\\hline
$u$ & $\hat u$ & $-3{\rm cos}^2{\theta_W}    {\rm sin}\theta_L   {\rm cos}{\theta_L}+3 {\rm sin}\theta_L   {\rm sin}^3{\theta_W}    {\rm cos}{\theta_L}   \epsilon$ & $-3 {\rm sin}{\theta_R}   {\rm sin}^2{\theta_W}    {\rm cos}{\theta_R}   \sqrt{{\rm cos}2\theta_W}\epsilon$\\
& & & $-3 {\rm sin}{\theta_R}   {\rm sin}^3{\theta_W}    {\rm cos}{\theta_R}   \epsilon$ \\\hline
$\hat u$ & $\hat u$ & $-3 {\rm cos}^2{\theta_W}    {\rm sin}^2\theta_L+4 {\rm cos}^2{\theta_W}    {\rm sin}^2{\theta_W}$ & $4 {\rm cos}^2{\theta_W}    {\rm sin}^2{\theta_W}+3 {\rm cos}^2{\theta_R}    {\rm sin}^2{\theta_W}    \sqrt{{\rm cos}2\theta_W}\epsilon$\\
& & $- (4-3 {\rm sin}^2\theta_L)   {\rm sin}^3{\theta_W}    \epsilon$ & $- (1+3 {\rm sin}^2{\theta_R})   {\rm sin}^3{\theta_W}    \epsilon$\\\hline\hline
\end{tabular}
\end{center}
\caption{\small{Analytical expressions for $A_{ff^\prime}^{Z}$ and $B_{ff^\prime}^{Z}$.}}
\label{tab:NCZ} }
\end{table}
$A_{ff^\prime}^Z,~B_{ff^\prime}^Z$ 
%and  $A_{ff^\prime}^{\hat Z},~B_{ff^\prime}^{\hat Z}$ 
are presented in Table~\ref{tab:NCZ}.
% and~\ref{tab:NCZp} respectively. 
%
The interactions of fermions with the SM Higgs and mirror Higgs are described in Eq.~\ref{eq:higgs}.
\begin{eqnarray}
{\cal L}_{S}=&&\frac{y_f}{\sqrt 2}\bar f \left[A_{ff^\prime}^{H}\frac{1-\gamma^5}{2}+B_{ff^\prime}^{H}\frac{1+\gamma^5}{2}\right]H f^{\prime}\nonumber\\
&&\frac{y_f}{\sqrt 2}\bar f \left[A_{ff^\prime}^{\hat H}\frac{1-\gamma^5}{2}+B_{ff^\prime}^{\hat H}\frac{1+\gamma^5}{2}\right]\hat H f^{\prime},
\label{eq:higgs}
\end{eqnarray}
where, $y_f$ is the Yukawa coupling of fermion $f$. The expressions for the coefficients $A_{ff^\prime}^{H},~B_{ff^\prime}^{H},~A_{ff^\prime}^{\hat H}~{\rm and}~B_{ff^\prime}^{\hat H}$ can be found in Table~\ref{tab:higgs}. It is important to note that in the limit $\epsilon=0$ and ${\rm cos}\theta_{L,R}=1$, the SM fermions decouple from the mirror fermions and we recover the SM 
%gauge boson 
couplings.

The decays of the TeV scale mirror fermions into $\hat W$, $\hat Z$ or $\hat H$ are kinematically forbidden since the mass of these mirror bosons are proportional to $\hat v \sim 10^7$ GeV. Because of the mixing of the mirror fermions with the ordinary fermions, the mirror fermions can decay into a SM fermion, and a $Z,~ W$ or a Higgs boson. The  {expressions for the partial} decay widths are:
\begin{eqnarray}
\Gamma({\hat f} \to f Z)&=&\frac{g^2}{36{\rm cos}^6\theta_w}\frac{\left({A_{ff^\prime}^{Z}}\right)^2+\left({B_{ff^\prime}^{Z}}\right)^2}{64\pi}\frac{M_{\hat f}^3}{M_Z^2}\left(1-\frac{M_Z^2}{M_{\hat f}^2}\right)^2\left( 1+2\frac{M_Z^2}{M_{\hat f}^2}\right),\nonumber\\
\Gamma({\hat f} \to f^\prime W)&=&\frac{g^2}{8}\frac{\left({A_{ff^\prime}^{W}}\right)^2+\left({B_{ff^\prime}^{W}}\right)^2}{16\pi}\frac{M_{\hat f}^3}{M_W^2}\left(1-\frac{M_W^2}{M_{\hat f}^2}\right)^2\left( 1+2\frac{M_W^2}{M_{\hat f}^2}\right),\nonumber\\
\Gamma({\hat f} \to f H)&=&\frac{y_f^2}{2}\frac{\left({A_{ff^\prime}^{H}}\right)^2+\left({B_{ff^\prime}^{H}}\right)^2}{64\pi}M_{\hat f}\left(1-\frac{M_H^2}{M_{\hat f}^2}\right)^2,
\end{eqnarray}  
where, $M_Z,~M_W,~M_H~{\rm and}~M_{\hat f}$ are the masses of $Z,~W$, Higgs and mirror fermion 
\begin{table}[h!]
\begin{center}
\begin{tabular}{||c|c||c|c||c|c||}
\hline \hline
$f$ & $f^\prime$ & $A_{ff^\prime}^{H}$ & $B_{ff^\prime}^{H}$ & $A_{ff^\prime}^{\hat H}$ & $B_{ff^\prime}^{\hat H}$  \\\hline\hline
$f$ & $f$ & ${\rm cos}\theta_L{\rm cos}\theta_R$ & ${\rm cos}\theta_L{\rm cos}\theta_R$ & ${\rm sin}\theta_L{\rm sin}\theta_R$ & ${\rm sin}\theta_L{\rm sin}\theta_R$\\\hline
$f$ & $\hat f$ & ${\rm sin}\theta_L{\rm cos}\theta_R$ & ${\rm cos}\theta_L{\rm sin}\theta_R$ & -${\rm cos}\theta_L{\rm sin}\theta_R$ & -${\rm sin}\theta_L{\rm cos}\theta_R$\\\hline
$\hat f$ & $\hat f$ & ${\rm sin}\theta_L{\rm sin}\theta_R$ & ${\rm sin}\theta_L{\rm sin}\theta_R$& ${\rm cos}\theta_L{\rm cos}\theta_R$ & ${\rm cos}\theta_L{\rm cos}\theta_R$\\\hline\hline
\end{tabular}
\end{center}
\caption{\small{Analytical expressions for $A_{ff^\prime}^{H}$, $B_{ff^\prime}^{H}$, $A_{ff^\prime}^{\hat H}$ and $B_{ff^\prime}^{\hat H}$.}}
\label{tab:higgs}
\end{table}
respectively. Apart from the known SM parameters and mirror fermion masses, the decay widths of mirror fermions depend on $\epsilon$, $\theta_L~{\rm and}~\theta_R$. For $\hat v \sim 10^7$ GeV, the value of $\epsilon$ is about $10^{-10}$. Therefore, the terms proportional to $\epsilon$ in the decay widths can be safely neglected. The mirror fermions decay widths depend primarily on the fermion mixing angles. According to Eq.~\ref{tan}, the fermion mixing angles are determined in terms of two parameters, namely, $\hat v$ and $M_{f\hat f}$.  
%------------------------------------------------------------------
\begin{figure}[h!]
\begin{center}
\epsfig{file=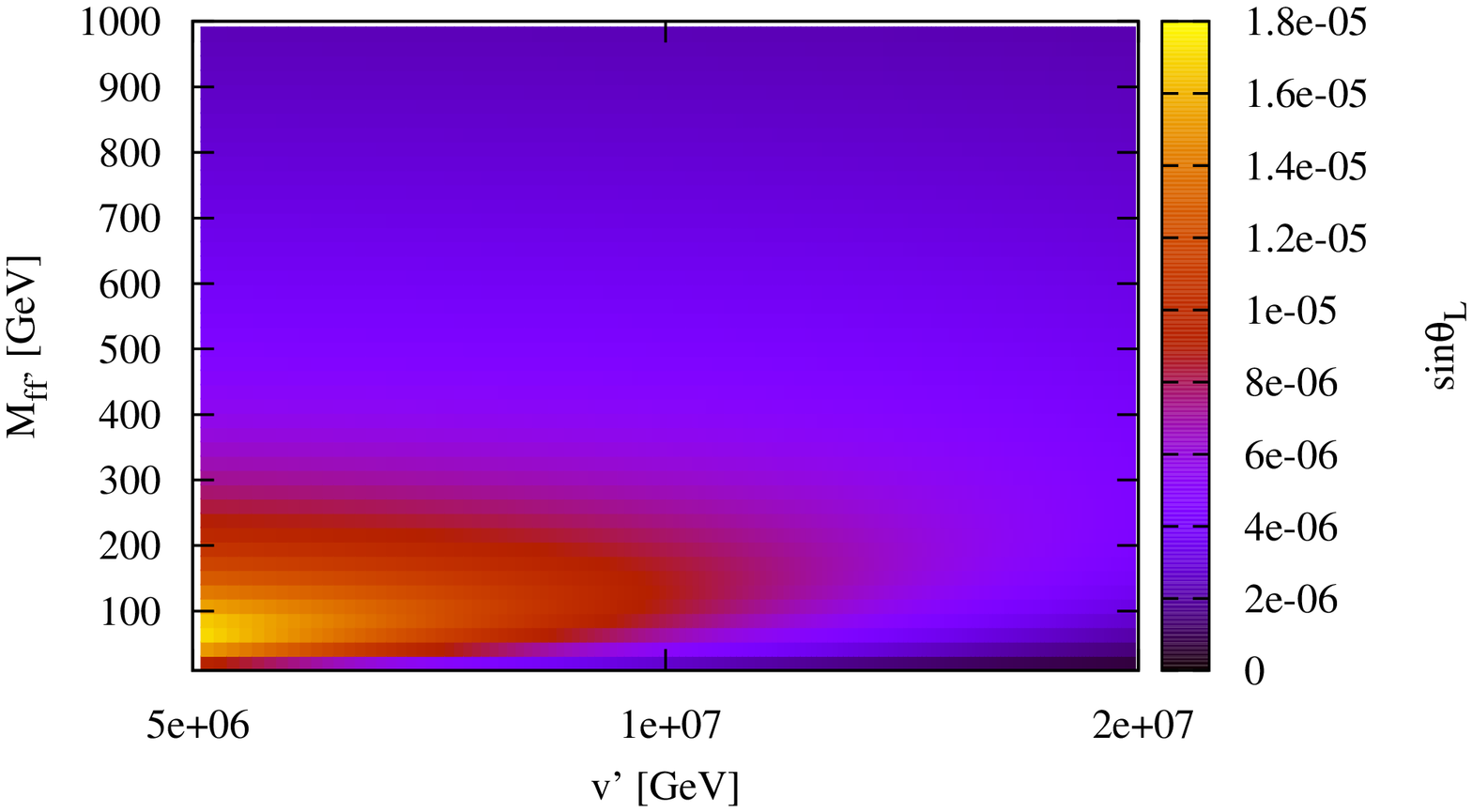,width=8cm,height=8cm,angle=-90}
\epsfig{file=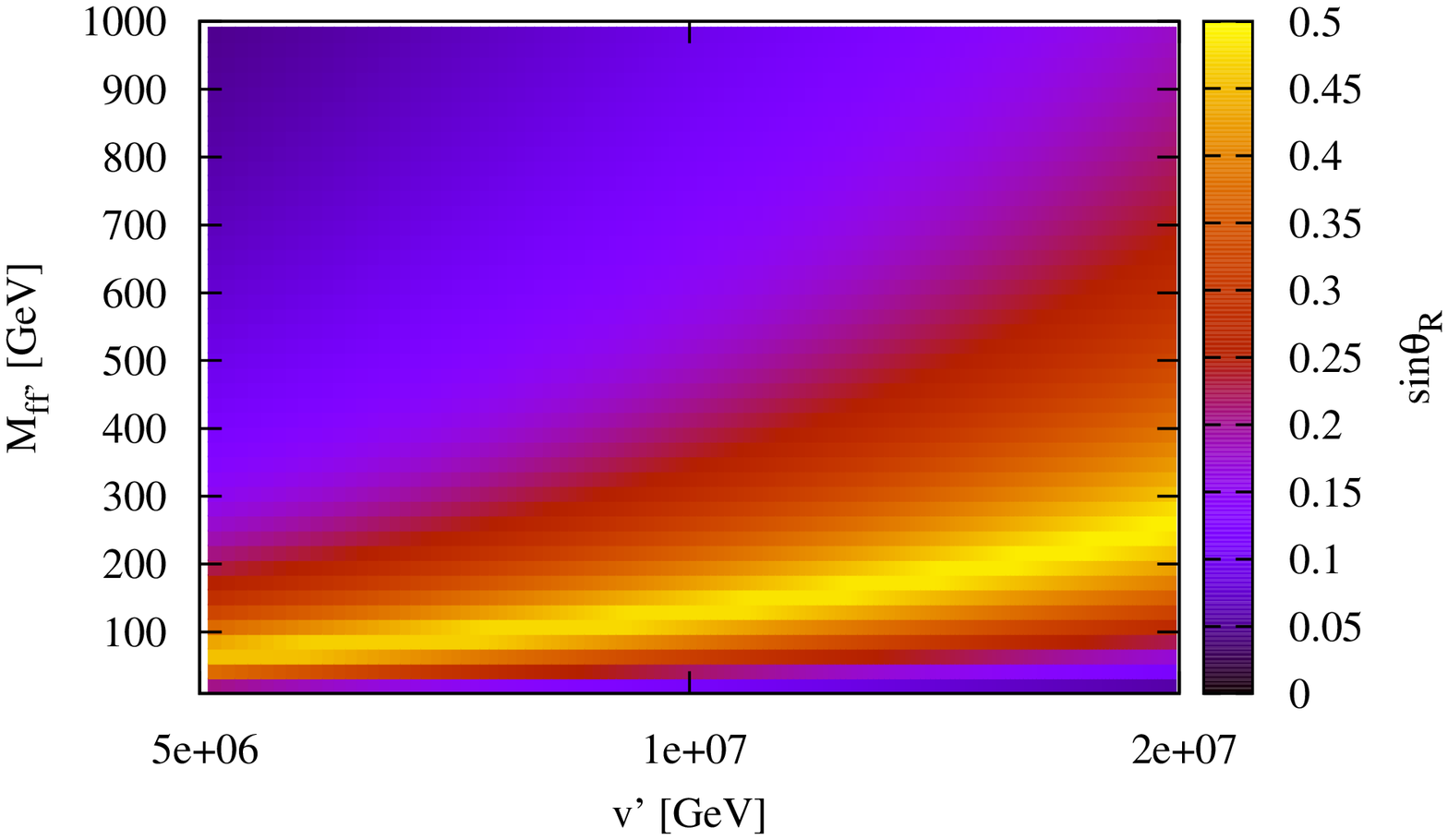,width=8cm,height=8cm,angle=-90}
\end{center}
\caption{\small{Fermion mixing angles, ${\rm sin}\theta_L$ (left panel) and ${\rm sin}\theta_R$ (right panel), for the up quark are presented by color gradient on the LRMM parameter space defined by $\hat v$ (along x-axis) and $M_{f\hat f}$ (along y-axis). The up quark Yukawa coupling, $y_u=1.3\times 10^{-5}$, and the SM VEV, $v=250$ GeV are assumed in these plots. }}
\label{mixing}
\end{figure}
%-------------------------------------------------------------------
Assuming the up quark Yukawa coupling, $y_u=1.3\times 10^{-5}$ and the SM VEV, $v=250$ GeV, in Fig.~\ref{mixing}, we show the mixing angles, ${\rm sin}\theta_L$ (left panel) and ${\rm sin}\theta_R$ (right panel), by color gradient, in the $\hat v$-$M_{f\hat f}$ plane. Eq.~\ref{tan}, shows that ${\rm tan}2\theta_L$ is suppressed by the SM quark mass ($\sim y_f v$) in the numerator and mirror 
quark mass ($\sim \sqrt{y_f^2 \hat v^2+2M_{f\hat f}^2}$) in the denominator. Therefore, for a MeV scale SM quark and TeV scale mirror partner, the value of ${\rm sin}\theta_L$ is about $10^{-6}$ which can be seen in Fig.~\ref{mixing} (left panel). Whereas, Fig.~\ref{mixing} (right panel) shows that,  ${\rm sin}\theta_R$ can be large depending on the values of $\hat v$ and $M_{f\hat f}$. 
%------------------------------------------------------------------
\begin{figure}[hb!]
\begin{center}
\epsfig{file=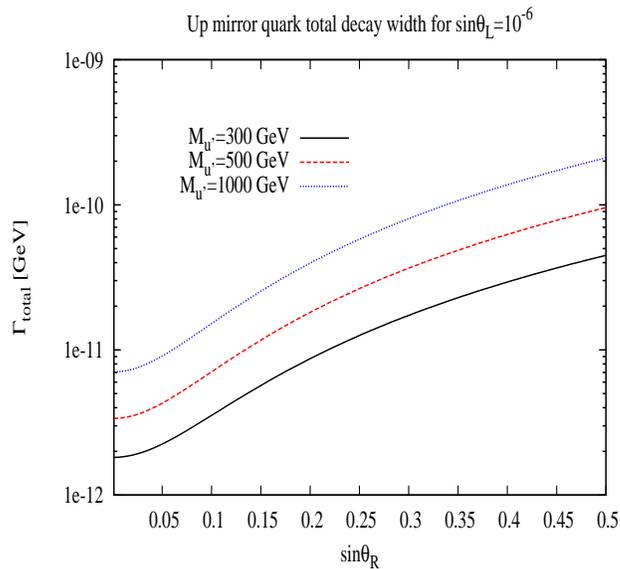,width=3.2in,height=3.3in,angle=-90}
\end{center}
\caption{\small{Total decay width of up-type mirror quark for three different values of 
$M_{\hat u}=300,~500~{\rm and}~1000$ GeV as a function of ${\rm sin}\theta_R$. We have 
assumed lowest possible value for ${\rm sin}\theta_L=10^{-6}$ in this plot.}}
\label{fig:width}
\end{figure}
%-------------------------------------------------------------------

The neutral (see Eq.~\ref{eq:NC}) and charge (see Eq.~\ref{eq:cc}) current interactions of mirror quarks with SM quarks and $Z/W$ bosons are suppressed by ${\rm sin}\theta_L$. Moreover, the interactions of mirror quarks with the SM quarks and Higgs boson are suppressed by the Yukawa couplings. Therefore, before going into the details of collider analysis, it is important to ensure that light mirror quarks decay inside the detectors of the LHC experiment. In Fig.~\ref{fig:width}, we plot the total decay width of up-type mirror 
quark as a function of ${\rm sin}\theta_R$ for three different values of the mirror quark mass, {\it viz.},
$M_{\hat u}=300,~500~{\rm and}~1000$ GeV. 
We have considered the lowest possible value of ${\rm sin}\theta_L=10^{-6}$ in Fig.~\ref{fig:width}. According to Fig.~\ref{fig:width}, the total decay width of up-type mirror quark is always greater than $10^{-12}$ GeV, $\Gamma_{total}>10^{-12}$ GeV, which  corresponds to a mean distance of $c\tau < 10^{-3}$ cm (without including Lorentz boost) traversed by a mirror quark inside a detector before its decay. These numbers assure us that 
the mirror quarks will always decay inside the detector for a wide range of model parameters.
%------------------------------------------------------------------
\begin{figure}[h!]
\begin{center}
\epsfig{file=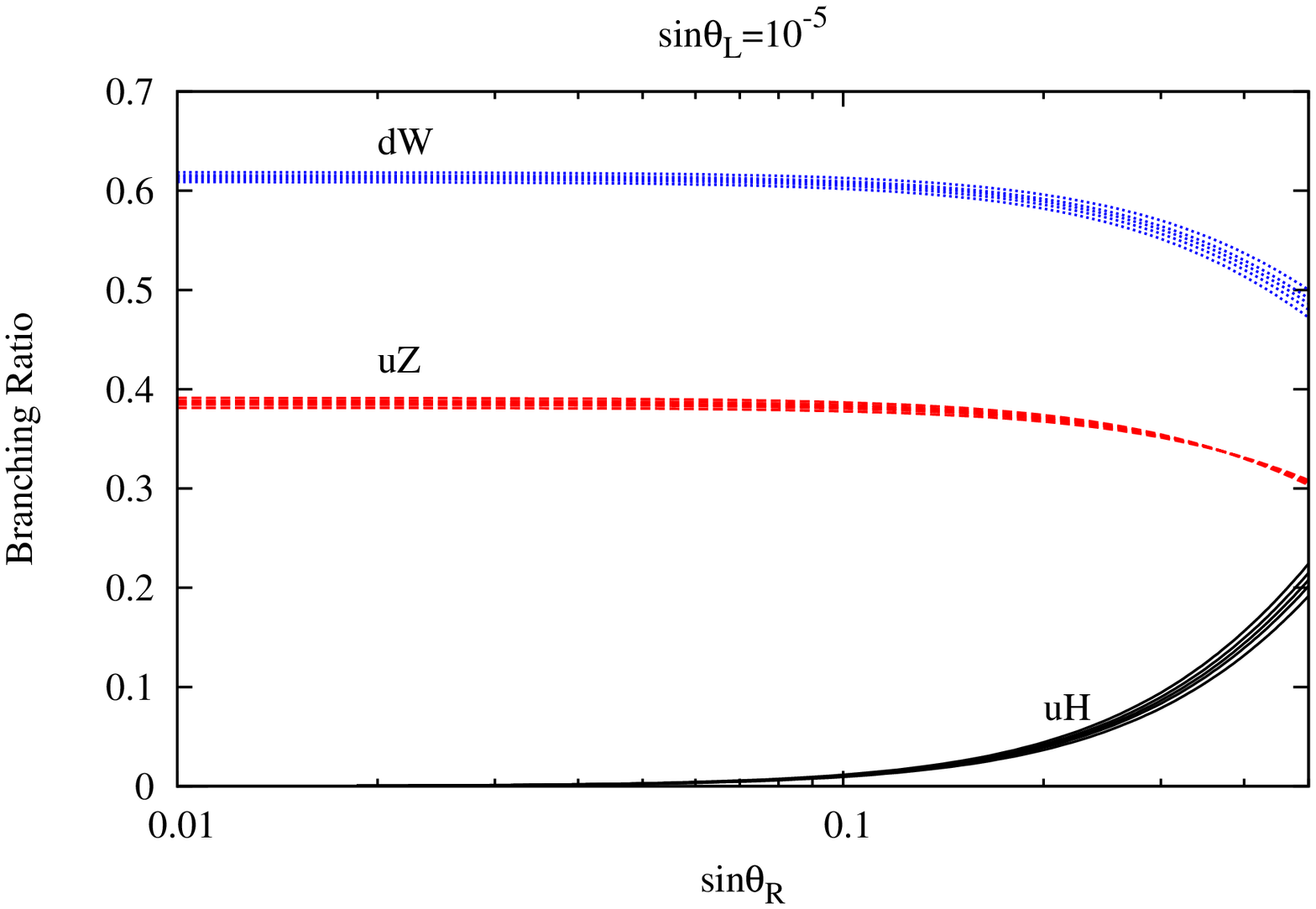,width=8cm,height=8cm,angle=-90}
\epsfig{file=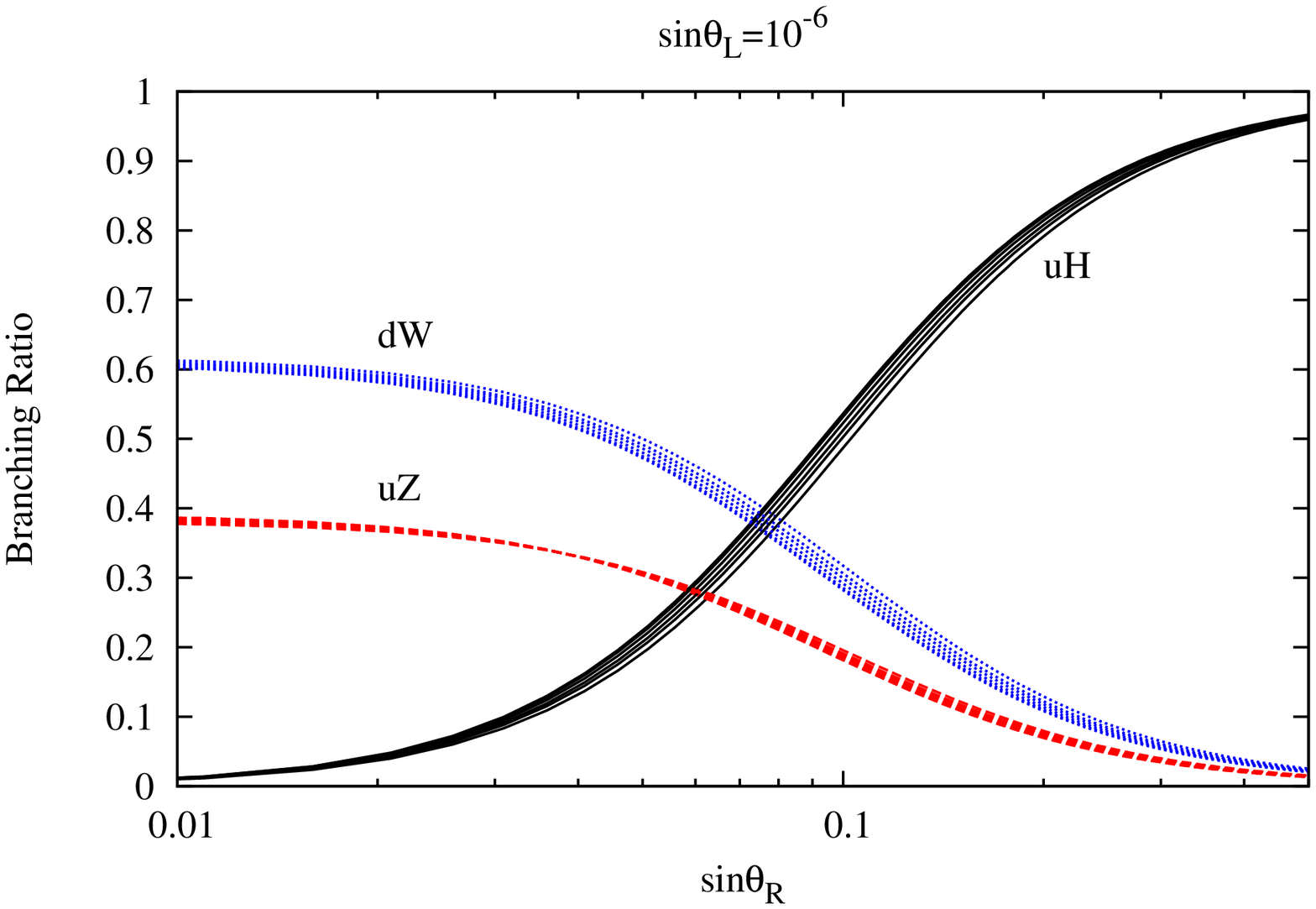,width=8cm,height=8cm,angle=-90}
\end{center}
\caption{\small{ Illustrating the up-type mirror quark branching ratios in $dW$, $uZ$ and $uH$ channel as a function of ${\rm sin}\theta_R$ for two different values of ${\rm sin}\theta_L=10^{-5}$ (left panel) and $10^{-6}$ (right panel). We have varied $\hat u$  mass over a range between 300 GeV to 1 TeV which gives rise to the bands instead of  lines.}}
\label{BR}
\end{figure}
%-------------------------------------------------------------------

In Fig.~\ref{BR}, we plot the branching ratios for the up-type mirror quark into $dW$, $uZ$ and $uH$ channel as a function of ${\rm sin}\theta_R$. We have assumed two different values of ${\rm sin}\theta_L=10^{-5}$ (left panel) and $10^{-6}$ (right panel). We have varied the mirror quark mass over 300 GeV to 1 TeV which gives rise to the bands in Fig.~\ref{BR}. Fig.~\ref{BR} (left panel) shows that for ${\rm sin}\theta_L=10^{-5}$, the decay of $\hat u$ into SM vector bosons dominates over the decay into Higgs boson. Whereas, for ${\rm sin}\theta_L=10^{-6}$ (right panel), the decay into vector bosons dominates only in the low ${\rm sin}\theta_R$ region (${\rm sin}\theta_R<0.08$).  

%%%%%%%
\subsection{Signature of mirror fermions at the LHC}
In this section, we will first discuss the production of TeV scale mirror quarks, namely ${\hat u} ~{\rm and}~{\hat d}$ quarks, at the LHC. As a consequence of the $Z_2$ symmetry, the couplings between a mirror quark and the SM particles are forbidden. Therefore, in presence of this $Z_2$ symmetry, the single production of the mirror fermions is not possible at the collider. As discussed in the previous section, spontaneous breaking of the $Z_2$ symmetry introduces mixing between the mirror and SM quarks and thus, gives rise to interactions between mirror and SM quarks with a $Z$, $W$ or Higgs boson. However, the single production rates of TeV scale mirror quarks via the $Z_2$ symmetry violating couplings are suppressed by the quark mixing angles. Therefore, in this work, we have considered the pair production of mirror quarks at the LHC. 

%------------------------------------------------------------------
\begin{figure}[h!]
\begin{center}
\epsfig{file=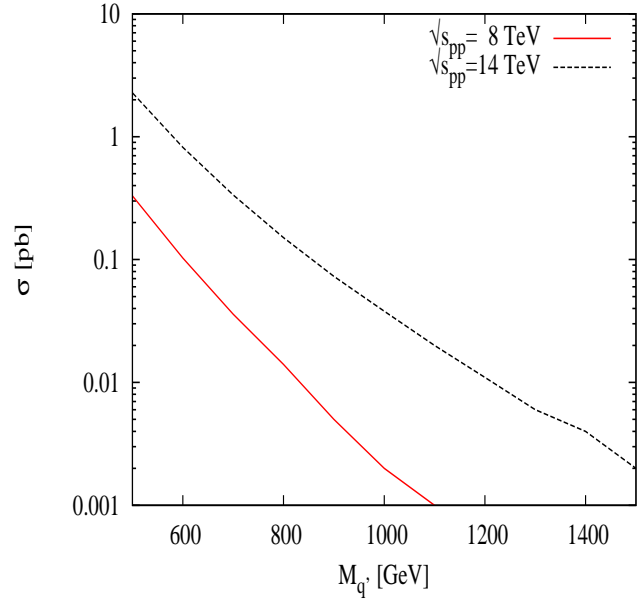,width=3.4in,height=3.5in,angle=-90}
\end{center}
\caption{\small{Pair production cross-sections of mirror quarks as a function of their
masses in proton proton collisions at center-of-mass energies 8 TeV and 14 TeV respectively.}}
\label{cross}
\end{figure}
\begin{figure}[h!]
\begin{center}
\epsfig{file=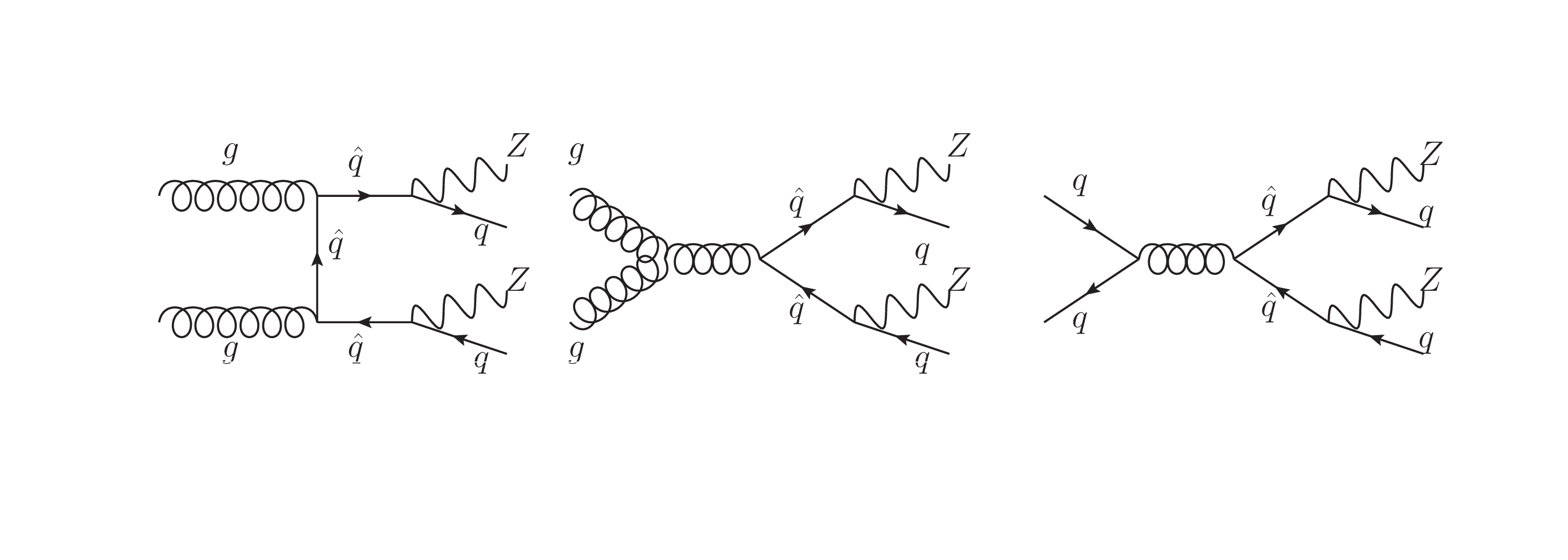,width=15cm,height=4cm,angle=0}
\end{center}
\caption{\small{Feynman diagrams for the ${\hat q} \bar {\hat q}$ production and their subsequent decay to $q Z$.} }
\label{FD}
\end{figure}

%-------------------------------------------------------------------
As the mirror quarks carry $SU(3)_C$ quantum numbers, they couple directly to the gluons. 
%All the mirror quarks have tree level couplings with respective mirror anti-quarks and a SM gluon. 
The pair production of TeV scale mirror quarks, namely ${\hat u} \bar {\hat u}$ and ${\hat d} \bar {\hat d}$ production, in a proton-proton collision therefore is analogous to that of the pair production of SM heavy quarks, the analytic expressions for which can be found in Ref.~\cite{top}. 
Both gluon-gluon ($gg$) and  
%(through $t(u)$-channel mirror quark exchange and $s$-channel SM gluon exchange) as well as from 
quark-antiquark ($q \bar q$) initial states 
%(through $s$-channel SM gluon exchange diagram) 
contribute to the pair production (${\hat q} \bar {\hat q}$) of mirror quarks (see Fig.~\ref{FD}). For numerical evaluation of the cross-sections, we have used a tree-level Monte-Carlo program incorporating CTEQ6L \cite{cteq6l} parton distribution functions. Both the renormalization and the
factorization scales have been set equal to the subprocess center-of-mass energy $\sqrt {\hat s}$. The ensuing leading-order (LO) ${\hat q} \bar {\hat q}$ production cross-sections are presented in Fig.~\ref{cross} as a function of mirror quark mass ($M_{\hat q}$) for two different values of the proton-proton center-of-mass energy {\it viz.,} $\sqrt{s}_{pp}=8$ TeV and $14$ TeV. While the NLO and NLL corrections can be well estimated by a proper rescaling of the corresponding results for $t\bar t$ production, we deliberately resist from doing so. With the K-factor expected to be large \cite{ttk}, our results would thus be a conservative one. The pair production cross section is found to be a few hundred femtobarns (fb) for mirror quark mass 
of close to a TeV. As discussed before, these mirror quarks once produced will decay within the detector.  
 %After a very short discussion about the pair production cross-sections of TeV scale mirror quarks,
 We now analyze the possible signatures of mirror quarks at the LHC following its decay properties.  Mirror 
quarks can decay into a $Z$-boson, a $W$-boson or Higgs boson in association with a SM quark: ${\hat q} \to q Z,~q^\prime W~{\rm and}~qH$. Thus the pair production of mirror quarks, at the LHC, gives rise to a pair of heavy SM bosons ($Z$-boson, $W$-boson or Higgs boson) in association with multiple jets in the final state. 
%------------------------------------------------------------------
%-------------------------------------------------------------------
In this work, we have focused on the signal with the vector bosons in the final states. 
%We do not consider the final states with Higgs bosons. Therefore, our collider analysis is only applicable to those part of 
We choose the LRMM parameter space where the decay of mirror quarks into vector bosons dominates over its decay into Higgs boson. Fig.~\ref{BR} shows that for negligible $\hat q \to qH$ branching ratio, the mirror quarks decay into $qW$ and $qZ$ pairs with about 61\% and 39\% branching probability respectively. In the rest of our analysis, we have used the above mentioned values for the decay probability to compute the signal cross-sections. Pair production and the decay of mirror quarks in to $qW$ and $qZ$ channels gives rise to the following signatures:
\begin{itemize}
\item {\it {2 jets}+2 $Z$ } final state arises when both mirror quarks decay into $qZ$ pairs.
 $$
pp \to {\hat q} \bar {\hat q} \to (qZ)(\bar q Z)
$$
The production and decay of mirror quarks in this channel are schematically shown in Fig.~\ref{FD}. 
\item {\it {2 jets}+$Z$+$W$ } final state results when one mirror quark decays into $qZ$ channel and other one decays into $qW$ channel.
$$
pp \to {\hat q} \bar {\hat q} \to (qZ)(\bar q^\prime W)
$$
\item If both mirror quarks decay into $qW$ channel then pair production of mirror quarks gives rise to {\it {2 jets}+2 $W$} final state. 
\end{itemize}  
We consider the reconstruction of mirror quark mass from the invariant mass distribution of $qZ$ pairs which is possible for the first two signal topologies only. Therefore, we have only considered  {\it {2 jets}+2 $Z$} and {\it {2 jets}+$Z$+$W$} final states for further 
analysis. Note that in the leptonic channel the $Z$ reconstruction would be very clean while for the {\it {2 jets}+$Z$+$W$}, even the $W$ can be 
reconstructed well as there is only a single neutrino in the final state. The $W$'s can be reconstructed in the all hadronic mode but with significant
challenge in efficiencies in a hadronic machine such as the LHC. So we have chosen to neglect the {\it {2 jets}+2 $W$} final state in our analysis.

\subsubsection{{\it {2 jets}+2 $Z$-bosons} signature}
In this section, we have investigated {\it 2 jets + 2 $Z$ } final state as a signature of mirror quarks in the framework of LRMM. We have used a parton level Monte-Carlo simulation to evaluate the cross-sections and different kinematic distributions for the signal. We have assumed that $Z$-bosons decaying into leptons (electrons and muons) can be identified at the LHC with good efficiency. Therefore, in our parton level analysis, we consider $Z$-boson as a standard object\footnote{All the cross-sections (signal as well as background) presented in the next part of this article are multiplied by the leptonic branching fraction (6.7\% in electron and muon channel) of the $Z$-boson.} without simulating its decay to leptons.  {We must however point out that the total number of signal events are crucial 
in identifying the $Z$ boson in the leptonic channel because of the small branching probability of the $Z$ decaying to charged leptons.}

The dominant SM background to the signal comes from the pair production of $Z$-bosons in association with two jets. Before going into the details of signal and background, it is important to list a set of basic requirements for jets to be visible at the detector. To parametrize detector acceptance and enhance signal to background ratio, we have imposed kinematic cuts ({\it Acc. Cuts}), listed in Table~\ref{cuts}, on the jets (denoted by $j_1$ and $j_2$) after ordering the jets according to their transverse momentum ($p_T$) hardness ($p_T^{j_1}>p_T^{j_2}$). It should also be realized that any detector has only a finite resolution. For a realistic detector, this applies to both energy/transverse momentum measurements as well as determination of the angle of motion. For our purpose, the latter can be safely neglected\footnote {The angular resolution is, generically, far superior to the energy/momentum resolutions and too fine to be of any consequence at the level of sophistication of this analysis.} and we simulate the former by smearing the jet energy with Gaussian functions defined by an energy-dependent width, $\sigma_E$:
\begin{equation}
\frac{\sigma_E}{E}=\frac{0.80}{\sqrt E}\oplus 0.05,
\label{gau}
\end{equation}
where, $\oplus$ denotes a sum in quadrature.

\begin{table}[h]
\begin{center}
\begin{tabular}{c|c|c}
\hline \hline
Kinematic Variable & Minimum value & Maximum value \\\hline\hline
$p_T^{j_1,j_2}$ & 100 GeV & - \\
$\eta^{j_1,j_2}$ & -2.5 & 2.5 \\
$\Delta R(j_1,j_2)$  & 0.7 & - \\\hline\hline
\end{tabular}
\end{center}
\caption{\small{Acceptance cuts on the kinematical variables. $p_T^{j_1,j_2}$ is the transverse momentum and $\eta^{j_1,j_2}$ is the rapidity of the jets. $\Delta R(j_1,j_2)=\sqrt{(\Delta \eta)^2+(\Delta \phi)^2}$ is the distance among the jets in the $\eta-\phi$ plane, with $\phi$ being the azimuthal angle.}}

\label{cuts}
\end{table}

The signal jets arise from the decay of a significantly heavy mirror quark to a SM $Z$ and jet. Due to the large phase space available for the decay of the mirror quarks, the resulting jets will be predominantly hard. Therefore, the large jet $p_T$ cuts, listed in Table~\ref{cuts}, are mainly aimed to reduce the SM background contributions. With the set of acceptance cuts (see Table~\ref{cuts}) and detector resolution defined in the previous paragraph, we compute the signal and background cross-sections at the LHC operating with $\sqrt s$ = 8 TeV and 14 TeV respectively and display them in Table~\ref{cs}. Table~\ref{cs} shows that signal cross-sections are larger than the background for lower values of mirror quark masses. However, if we increase $M_{\hat q}$, signal cross-sections fall sharply as the pair production cross section for the mirror quarks fall with increasing mass.
\begin{table}[h]
\begin{center}
\begin{tabular}{||c|c|c||c|c||c|c|c||c|c||}
\hline \hline
\multicolumn{5}{||c||}{$\sqrt s$= 8 TeV} & \multicolumn{5}{|c||}{$\sqrt s$= 14 TeV}\\
\multicolumn{5}{||c||}{Cross-sections in fb} & \multicolumn{5}{|c||}{Cross-sections in fb}\\\hline\hline
\multicolumn{3}{||c||}{Signal} & \multicolumn{2}{|c||}{Background} & \multicolumn{3}{|c||}{Signal} &  \multicolumn{2}{|c||}{Background}\\\hline
%\cline{1-2}\cline{4-5}
$M_{\hat q}$ [GeV] & {\it A.C.} & {\it S.C.} & {\it A.C.} & {\it S.C.} & $M_{\hat q}$ [GeV]&  {\it A.C.} & {\it S.C.} & {\it A.C.} & {\it S.C.}\\\hline\hline
300 & 1.65 & 1.07 &       & 0.08 &   400 & 2.93 & 1.5  &      & 0.22 \\
350 & 0.92 & 0.52 & 0.35  & 0.07 &   500 & 1.04 & 0.48 & 1.36 & 0.14 \\
400 & 0.5  & 0.26 &       & 0.05 &   600 & 0.40 & 0.18 &      & 0.09\\\hline\hline
\end{tabular}
\end{center}
\caption{\small{Signal and SM background cross-section after the acceptance cuts ({\it A.C.}) and selection cuts ({\it S.C.}) for two different values of proton-proton center-of-mass energies. Signal cross-sections ($\sigma_{Signal}$) are presented for three different values of mirror quark masses ($M_{\hat q}$).}}
\label{cs}
\end{table}

%------------------------------------------------------------------
\begin{figure}[t]
\begin{center}
\epsfig{file=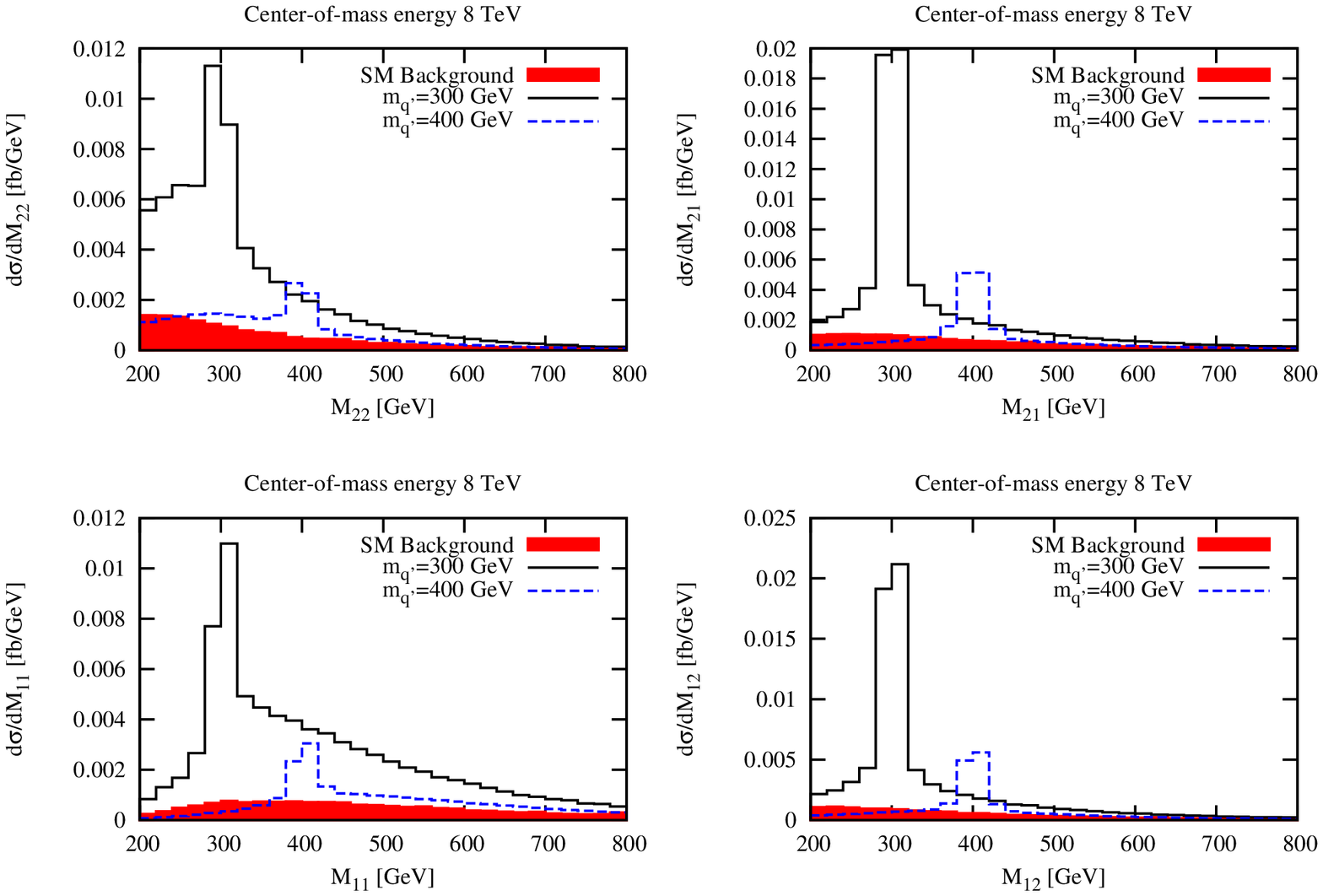,width=8cm,height=8cm,angle=-90}
\epsfig{file=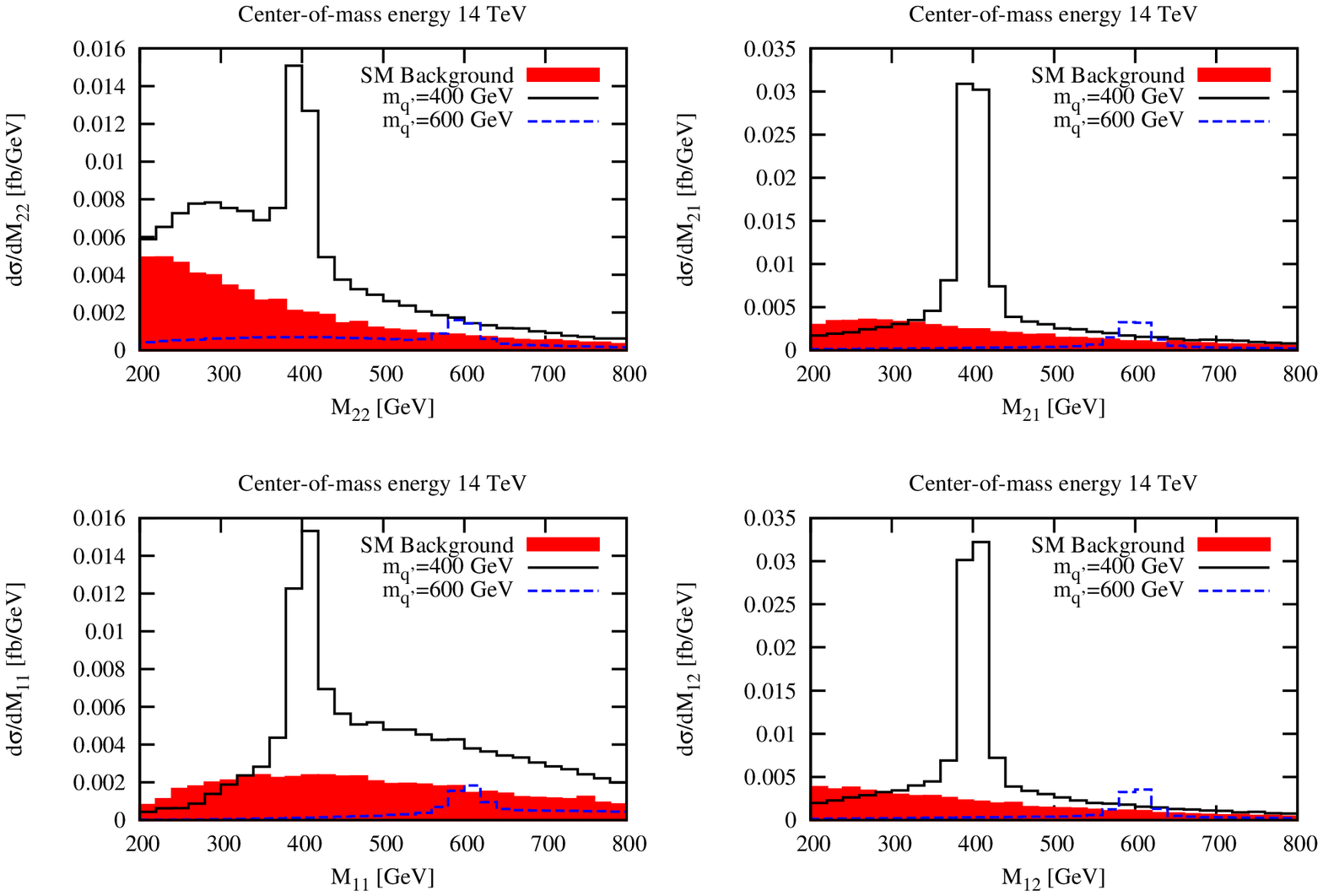,width=8cm,height=8cm,angle=-90}
\end{center}
\caption{\small{Jet-$Z$ invariant mass distributions after ordering the jets ($p_T^{j_1}>p_T^{j_2}$) and $Z$'s ($p_T^{Z_1}>p_T^{Z_2}$) according to their $p_T$ hardness for the LHC with center-of-mass energy 8 TeV (left panel) and 14 TeV (right panel).}}
\label{Invmass}
\end{figure}
%-------------------------------------------------------------------

Since the mirror quarks decay into a jet and $Z$-boson, the signal is characterized by a peak at $M_{\hat q}$ in the invariant mass distributions of jet-$Z$ pairs. The signal consists of two jets and two $Z$-bosons. In absence of any knowledge about the right jet-$Z$ pair arising from a particular $\hat q$ decay, we have ordered the jets and $Z$'s according to their $p_T$ hardness ($p_T^{j_1}>p_T^{j_2}$ and $p_T^{Z_1}>p_T^{Z_2}$) and constructed invariant mass distributions in the jet-$Z$ pairs as follows: $M_{11}=$ Invariant mass of $j_1~{\rm and}~Z_1$; $M_{12}=$ Invariant mass of $j_1~{\rm and}~Z_2$; $M_{21}=$ Invariant mass of $j_2~{\rm and}~Z_1$ and $M_{22}=$ Invariant mass of $j_2~{\rm and}~Z_2$. The four invariant mass distributions (for both signal and the SM background) are presented in Fig.~\ref{Invmass} for the LHC with center-of-mass energy 8 TeV (left panel) and 14 TeV (right panel). In Fig.~\ref{Invmass}, we have presented the signal invariant mass distributions for two different values of $M_{\hat q}$. We have included the leptonic branching ratio (6.7\% into electron and muon channel) of $Z$-boson into the cross-section in the Fig.~\ref{Invmass}. Fig.~\ref{Invmass} shows that the signal peaks are clearly visible over the SM background. Moreover, it is important to notice that signal peaks are more prominent in $M_{12}$ and $M_{21}$ distributions compared to $M_{11}$ and $M_{22}$ distributions. Due to the momentum conservation in the transverse direction at the LHC, both the mirror quarks are produced with equal and opposite transverse momentum\footnote{We do not consider initial/final state radiation (ISR/FSR) in our analysis. In presence of ISR/FSR, the transverse momentum of the mirror quarks might not be exactly equal and opposite.}. Therefore, if the decay of a particular mirror quark gives rise to the hardest jet then it is more likely that the $Z$-boson arising in the same decay will be the softest one. $M_{12}(M_{21})$ is the invariant mass of hardest-softest (softest-hardest) jet-$Z$ pairs which come from the decay of a particular $\hat q$ in most of the events. As a result, we observe more prominent peaks in the signal $M_{12} (M_{21})$ distribution compared to the $M_{11} (M_{22})$ distribution. In our analysis, we have utilized this feature of the signal for the further enhancement of signal to background ratio. Our final event selection criteria ({\it S.C.}) is summarized in the following:
\begin{itemize}
\item To ensure the observability of a peak for a given luminosity in the signal $M_{12}$ distribution, we have imposed the following criteria: (i) There are atleast 5 signal events in the peak bin. (ii) The number of signal events in the peak bin is greater than the $3\sigma$ fluctuation of SM background events in the same bin. 
\item If the signal peak in $M_{12}$ distribution is detectable then we selected events in the bins corresponding to the peak in the $M_{12}$ distribution and its four (two on the left hand side and two on the right hand side) adjacent bins as signal events. We have used a bin size of 20 GeV.
\item The total number of SM background events is given by the sum of events of the above mentioned five bins in the background $M_{12}$ distribution.  
\end{itemize}
After imposing the final event selection criteria, the signal and background cross-sections for different $M_{\hat q}$ and $\sqrt s$ are presented in Table~\ref{cs}. Table~\ref{cs} shows that selection cuts significantly suppress the SM background cross-section, whereas, signal cross-sections are reduced only by a factor $\sim$ 2.

\begin{figure}[t]
\begin{center}
\epsfig{file=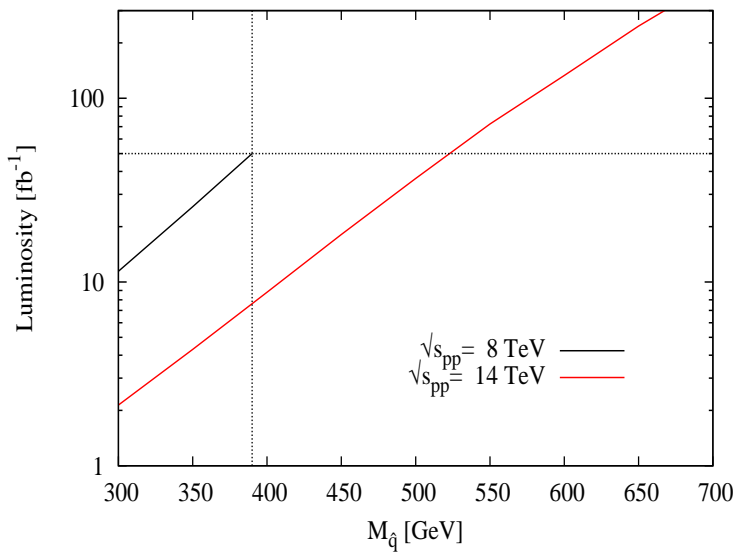,width=8cm,height=10cm,angle=-90}
\end{center}
\caption{\small{Required luminosity for $5\sigma$ discovery is plotted as a function of $M_{\hat q}$ for the LHC with center-of-mass energy 8 TeV and 14 TeV.} }
\label{lumi}
\end{figure}

After discussing the characteristics features of the signal and the SM background, we are now equipped enough to discuss the discovery reach of this scenario at the LHC with center-of-mass energy 8 TeV and 14 TeV. We define the signal to be observable for an integrated luminosity ${\cal L}$ if,
\begin{itemize}
\item
\begin{equation}
\frac{N_{S}}{\sqrt{N_B}} \ge 5 ~~~~~~ {\rm for}~~~~~ 0< N_B \le 5 N_S,
\label{discovery}
\end{equation}
where, $N_{S(B)}=\sigma_{S(B)} {\cal L}$, is the number of signal (background) events for an integrated 
luminosity ${\cal L}$.
\item For zero number of background event, the signal is observable if there are at 
least five signal events. 
\item In order to establish the discovery of a small signal (which could be 
statistically significant i.e. $N_S/\sqrt{N_B}\ge 5$) on top of a large background, we need to know the 
background with exquisite precision. However, such precise determination of the SM background is beyond 
the scope of this present article. Therefore, we impose the requirement $N_B\le 5 N_S$ to avoid such 
possibilities.
\end{itemize}
%\begin{equation}
%\sigma_S > \frac{N^2}{{\cal L}}\left[ 1+2\frac{\sqrt {{\cal L} \sigma_B}}{N}\right],
%\label{discovery}
%\end{equation} 
%where, $\sigma_S~{\rm and}~\sigma_B$ are the signal and background cross-sections, respectively and $N=2.5$ for $5\sigma$ discovery. 
The signal and background cross-sections in Table~\ref{cs} shows that at the LHC with center-of-mass energy 14 TeV, 500 GeV mirror quark mass can be probed with integrated luminosity 16 fb$^{-1}$. In Fig.~\ref{lumi}, we have presented the required luminosity for $5\sigma$ discovery as a function of $M_{\hat q}$ for the LHC with center-of-mass energy 8 TeV and 14 TeV.

%------------------------------------------------------------------
%-------------------------------------------------------------------

\subsubsection{{\it {Two jets}+$Z$-boson+$W$-boson} signature}
Another interesting final state results from the pair production of mirror quarks which then decay to give {\it {2 jets}+$Z$+$W$} signal. This happens when one mirror quark decays into $qZ$ while the other one decays into $qW$. As before we have considered the $Z$ boson as a standard object without simulating its decay to leptons (electrons and muons). 
%However, in the hadron collider environment, reconstruction of $W$-bosons in the leptonic decay channel ($W^\pm \to l^\pm \nu$) is difficult because of the unknown longitudinal boost of the center-of-mass frame. Therefore, $W$-bosons can not be treated as standard object like $Z$-bosons.
Even the $W$ boson can be reconstructed to a certain efficiency in the leptonic channel, where the neutrino $p_z$ is determined by using the $W$ mass constraints. This is possible because of a single neutrino in the final state. However, we have chosen to ignore the $W$ as a standard object since the $qZ$ resonance will be much more 
well defined and with less ambiguity.
In our parton level Monte-Carlo analysis, we have simulated the decay of $W$ bosons into leptons (electron and muons only) and neutrinos. Electrons and muons show charge tracks in the tracker and are detected at the electromagnetic calorimeter and muon detector respectively. However, neutrinos remain invisible in the detector and give rise to a imbalance in the visible transverse momentum vector which is known as missing transverse momentum ($\slashed{p}_T$). Therefore, the resulting signature in this case will be {\it {2 jets}+1 charged lepton + $Z$ + $\slashed{p}_T$}. 
\begin{table}[t!]
\begin{center}
\begin{tabular}{c|c|c}
\hline \hline
Kinematic Variable & Minimum value & Maximum value \\\hline\hline
$p_T^{l}$ & 25 GeV & - \\
$\eta^{l}$ & -2.5 & 2.5 \\
$\Delta R(l,j_{1,2})$  & 0.4 & - \\\hline\hline
\end{tabular}
\end{center}
\caption{\small{Acceptance cuts on the kinematical variables. $p_T^{l}$ is the transverse momentum and $\eta^{l}$ is the rapidity of the lepton. $\Delta R(l,j_{1,2})$ is the distance among the jet-lepton pairs in the $\eta-\phi$ plane, with $\phi$ being the azimuthal angle.}}
\label{cut_L}
\end{table}
%-----------------------------

The dominant SM background to the signal arises from the production of $ZW$ pairs in association with two jets. Both signal and background jets energy are smeared by a Gaussian function defined in Eq.~\ref{gau}. To ensure the visibility of the jets at the detector, acceptance cuts listed in Table~\ref{cuts} are applied on the jets. The 
%-----------------------------------------------
 \begin{table}[h!]
\begin{center}
\begin{tabular}{||c|c|c|c||c|c|c||c|c|c|c||c|c|c||}
\hline \hline
\multicolumn{7}{||c||}{$\sqrt s$= 8 TeV} & \multicolumn{7}{|c||}{$\sqrt s$= 14 TeV}\\
\multicolumn{7}{||c||}{Cross-sections in fb} & \multicolumn{7}{|c||}{Cross-sections in fb}\\\hline\hline
\multicolumn{4}{||c||}{Signal} & \multicolumn{3}{|c||}{Background} & \multicolumn{4}{|c||}{Signal} &  \multicolumn{3}{|c||}{Background}\\\hline
%\cline{1-2}\cline{4-5}
$M_{\hat q}$ & {\it A.C.} & {\it Cut} & {\it Cut} & {\it A.C.} & {\it Cut} & {\it Cut}  & $M_{\hat q}$ &  {\it A.C.} &  {\it Cut} & {\it Cut}  & {\it A.C.} & {\it Cut} & {\it Cut}\\
GeV & & {\it I}& {\it II} & & {\it I}& {\it II} &GeV & & {\it I}& {\it II} & & {\it I}& {\it II}\\\hline\hline
300 & 14.4 & 7.28 & 3.13 &       &      & 0.74 & 400 & 26.3 & 18.1 & 6.46 & & & 2.13\\
350 & 8.01  & 4.85 & 1.92 & 6.69 & 2.81 & 0.63 & 500 & 9.36  &  7.33 & 2.39 & 26.4 & 11.9 & 1.39\\
400 & 4.35 & 2.98 & 1.11 & & & 0.51 & 600 & 3.6 & 3.07 & 0.95 & & & 0.90 \\\hline\hline
\end{tabular}
\end{center}
\caption{\small{Signal and SM background cross-section after the acceptance cuts ({\it A.C.}), {\it Cut I} and {\it Cut II} for two different values of proton-proton center-of-mass energies. Signal cross-sections ($\sigma_{Signal}$) are presented for three different values of mirror quark masses ($M_{\hat q}$).}}
\label{cs_ZW}
\end{table}
%--------------------------------------
acceptance cuts for the lepton are listed in Table~\ref{cut_L}. We do not apply any cuts on the missing transverse momentum. With these set of cuts ({\it A.C.}) on jets (see Table~\ref{cuts}) and lepton (see Table~\ref{cut_L}), we have computed the signal and background cross-sections for the LHC with 8 TeV and 14 TeV center-of-mass energy and presented in Table~\ref{cs_ZW}. Table~\ref{cs_ZW} shows that for relatively large mirror quark masses, signal cross-sections are much smaller than the SM background cross-section. For example, at the LHC with 14 TeV center-of-mass energy, the signal to background ratio is 0.14 after acceptance cuts for $m_{\hat q}=600$ GeV. 

\begin{figure}[t]
\begin{center}
\epsfig{file=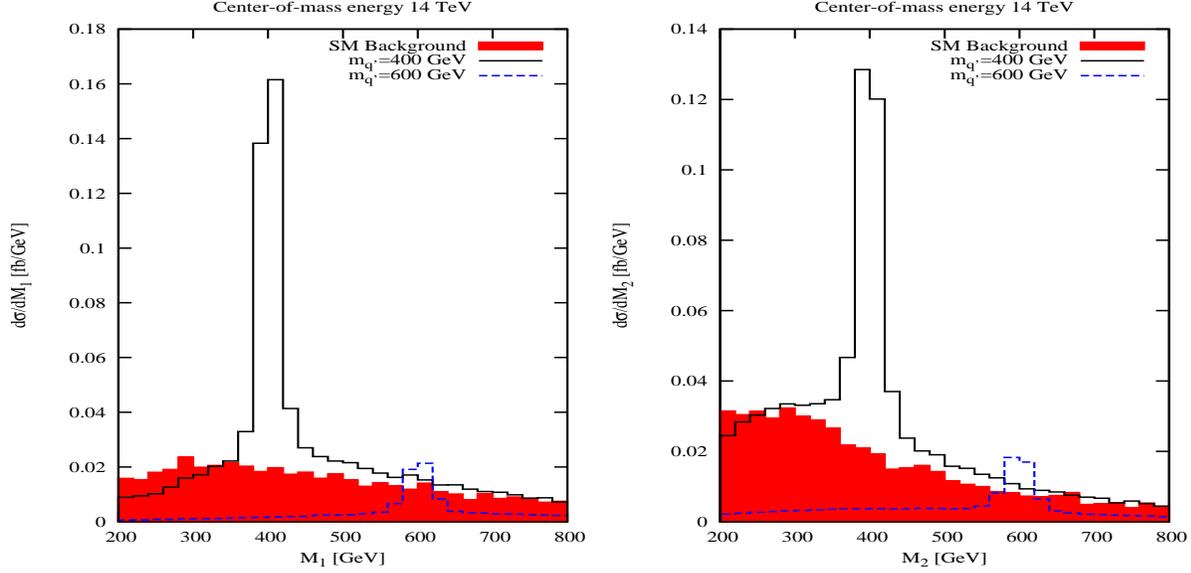,width=8cm,height=16cm,angle=-90}
\end{center}
\caption{\small{Jet-$Z$ invariant mass distributions after {\it Cut I} for the signal ($m_{\hat q}=400~{\rm and}~600$ GeV) and the SM background at the LHC with 14 TeV center-of-mass energy.}}
\label{inv_WZ}
\end{figure}

The signal contains a lepton and  $\slashed{p}_T$ arises from the decay of a $W$-boson. The SM background  lepton and  $\slashed{p}_T$ also results from the $W$-boson decay. However, the signal $W$-boson will be boosted in most of the events since it arises from the decay of a TeV scale mirror quark. We have tried to exploit this feature of the signal for the further enhancement of signal to background ratio. We have examined the following kinematic distributions:
\begin{itemize}
\item In Fig.~\ref{ptlep}, we have presented normalized lepton $p_T$ (left panel) and missing $p_T$ (right panel) distributions for the signal ($m_{\hat q}=400~{\rm and}~600$ GeV) and the SM background at the LHC with $\sqrt s=14$ TeV. The boost of the signal $W$-boson results into a long tail in the signal lepton and missing $p_T$ distributions. Fig.~\ref{ptlep} shows that harder cuts on the lepton and/or missing $p_T$ will suppress the SM background significantly. However, these cuts will also reduce signal cross-sections considerably. For example,  
a kinematic requirement of $p_T> 75$ GeV on the charged lepton at 14 TeV LHC will reduce 45\% of the SM background and 25\% of the signal for $m_{\hat q}=600$ GeV. As a result, we do not use any further cuts on 
lepton and/or missing $p_T$.

%------------------------------------------------------------------
\begin{figure}[ht!]
\begin{center}
\epsfig{file=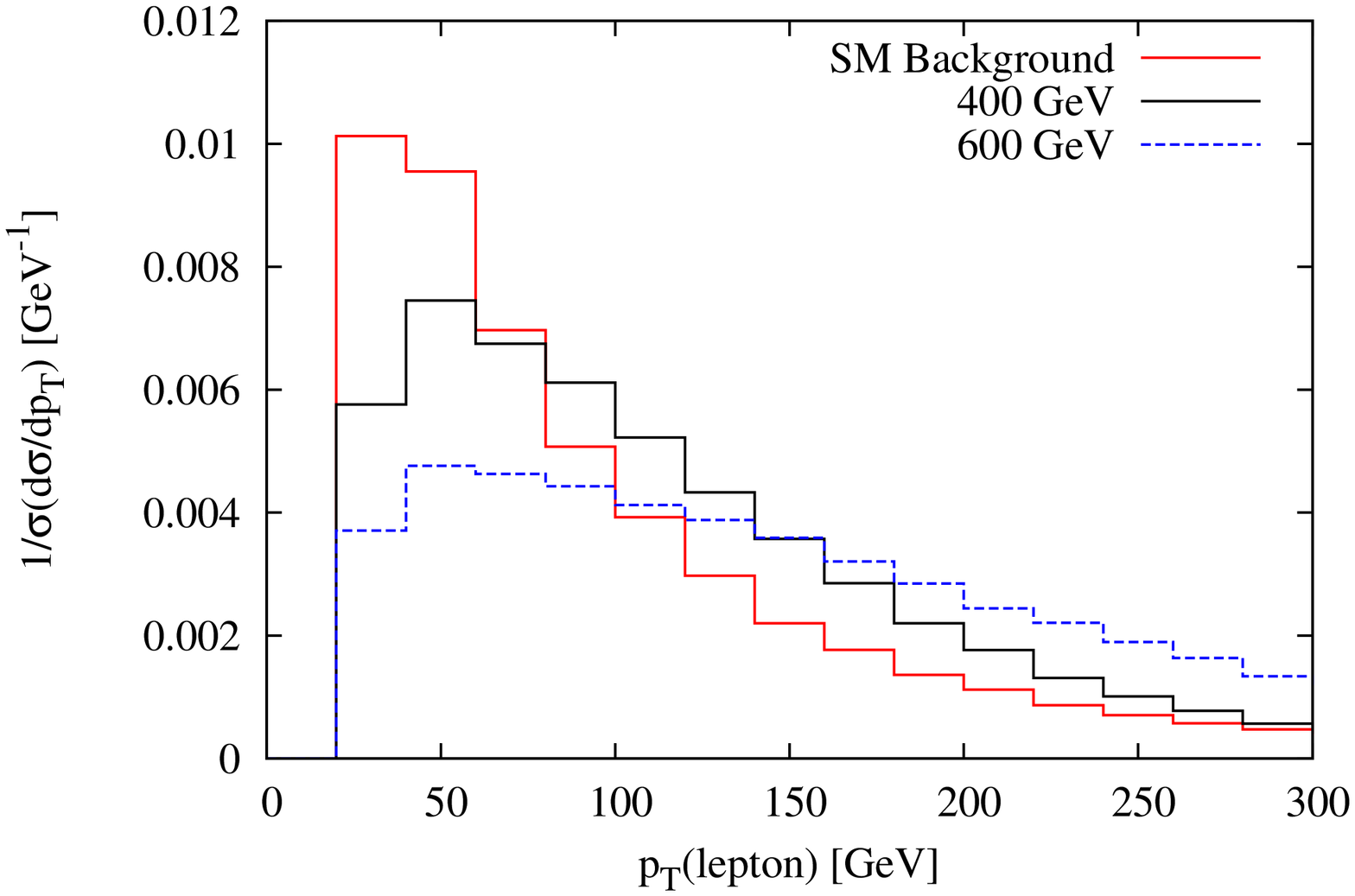,width=8cm,height=8cm,angle=-90}
\epsfig{file=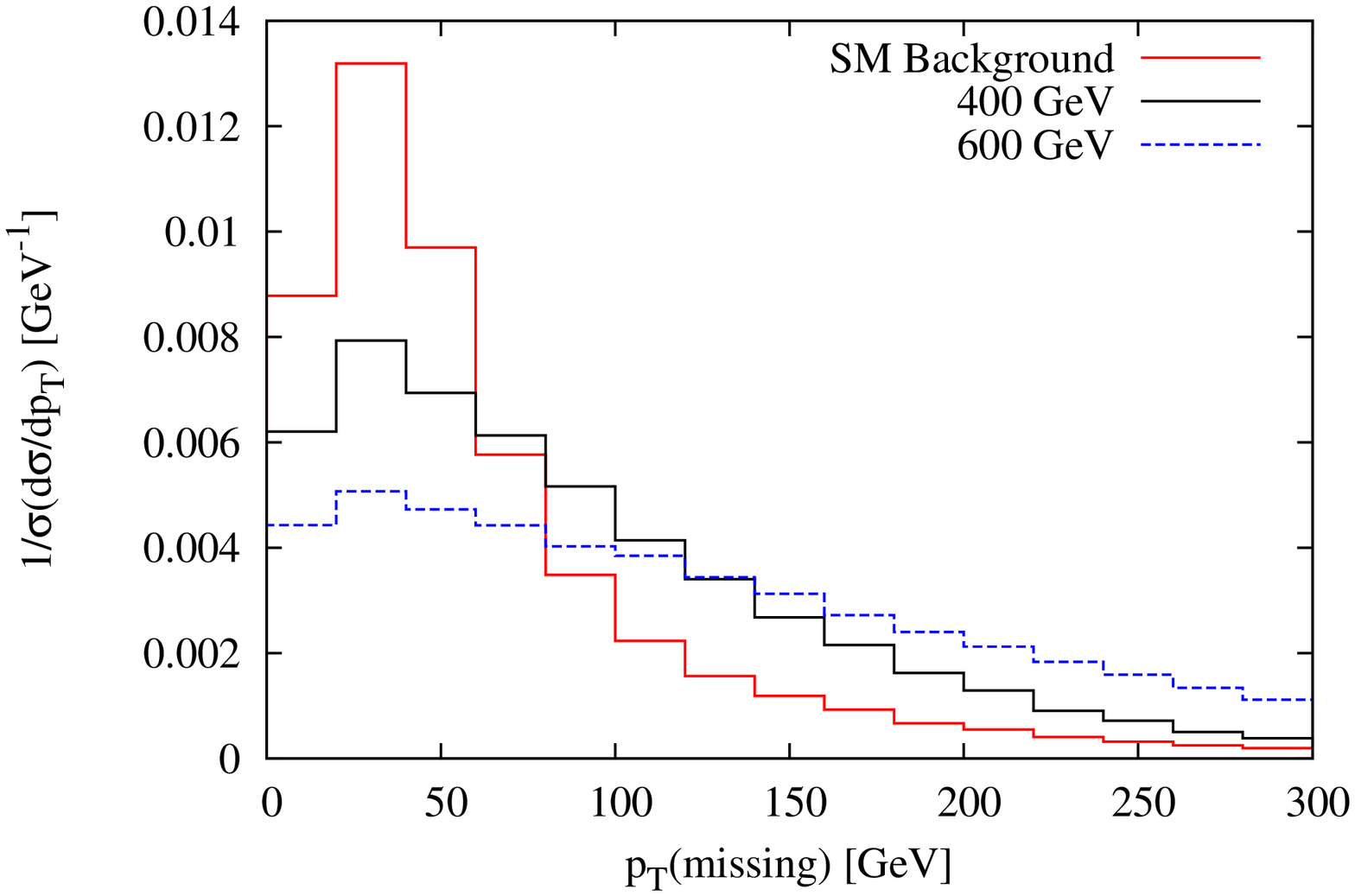,width=8cm,height=8cm,angle=-90}
\end{center}
\caption{\small{Normalized lepton $p_T$ (left panel) and missing $p_T$ (right panel) distributions for the signal ($m_{\hat q}=400~{\rm and}~600$ GeV) and the SM background after the acceptance cuts at the LHC with $\sqrt s=14$ TeV.} }
\label{ptlep}
\end{figure}
%-------------------------------------------------------------------
\item Since the signal $W$-boson is boosted, we expect that the signal lepton and neutrino will be collimated. Therefore, it is viable to study the azimuthal angle ($\Delta \phi$) between lepton transverse momentum vector ($\vec p_T^{l}$) and missing transverse momentum vector ($\vec{\slashed{p}}_T$). In Fig.~\ref{angle}, we have presented normalized $\Delta \phi(\vec p_T^{l},\vec{\slashed{p}}_T)$ distributions for the signal ($m_{\hat q}=400~{\rm and}~600$ GeV) and the SM background  at the LHC with $\sqrt s=14$ TeV. Since the background $W$-bosons are predominantly produced with small transverse momentum, background $\Delta \phi(\vec p_T^{l},\vec{\slashed{p}}_T)$ distribution is almost flat (see Fig.~\ref{angle}). Whereas, the signal $\Delta \phi(\vec p_T^{l},\vec{\slashed{p}}_T)$ distributions peaks in the small $\Delta \phi(\vec p_T^{l},\vec{\slashed{p}}_T)$ region. As a result, we have imposed an upper bound of 1 on the azimuthal angle between lepton $p_T$ vector and missing $p_T$ vector: $\Delta \phi(\vec p_T^{l},\vec{\slashed{p}}_T) < 1$. We collectively call acceptance cuts and $\Delta \phi(\vec p_T^{l},\vec{\slashed{p}}_T) < 1$ cut as {\it Cut I}. The signal and background cross-sections after {\it Cut I} are presented in Table~\ref{cs_ZW}. For 14 TeV center-of-mass energy, $\Delta \phi(\vec p_T^{l},\vec{\slashed{p}}_T) < 1$ cut reduces 55\% of the SM background and 14\% of the signal for $m_{\hat q}=600$ GeV and thus, enhances the signal to background ratio by a factor about 2. 
%------------------------------------------------------------------
\begin{figure}[h!]
\begin{center}
\epsfig{file=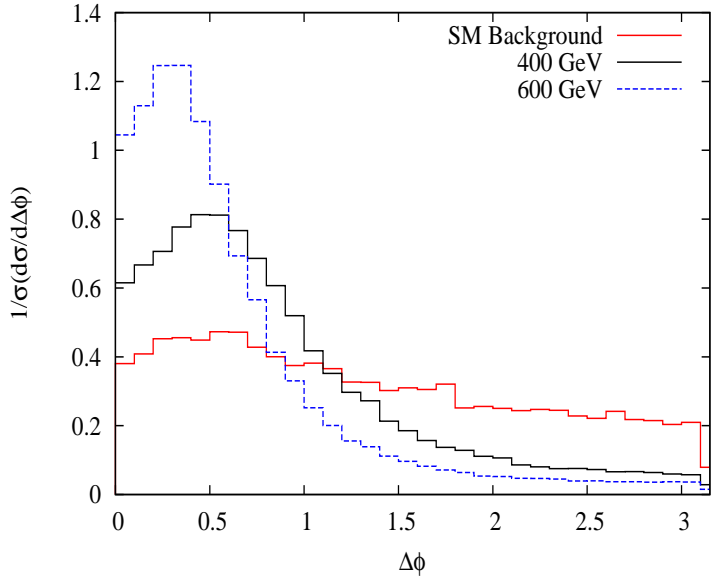,width=8cm,height=10cm,angle=-90}
\end{center}
\caption{\small{Normalized azimuthal angle $\Delta \phi(\vec p_T^{l},\vec{\slashed{p}}_T)$  distributions between lepton $p_T$ vector and missing $p_T$ vector after the acceptance cuts at the LHC with $\sqrt s=14$ TeV. Signal distributions are presented for two different values of mirror quark mass ($m_{\hat q}=400~{\rm and}~600$ GeV).}}
\label{angle}
\end{figure}
%-------------------------------------------------------------------

\item After $\hat q \bar{\hat q}$ production, one mirror quark decays into $qZ$ pair. Therefore, signal jet-$Z$ invariant mass distribution is characterized by a peak at $m_{\hat q}$. After ordering the jets according to their $p_T$ hardness ($p_T^{j_1}>p_T^{j_2}$), we have constructed two invariant mass: (i) $M_1$: invariant mass of $j_1$-$Z$ pair and (ii) $M_2$: invariant mass of $j_2$-$Z$ pair. The signal and background invariant mass distributions are presented in Fig.~\ref{inv_WZ} for the LHC with $\sqrt s=14$ TeV. For the further enhancement of signal to background ratio, we have imposed cuts on $M_2$ in a way similar to that discussed in the previous section. This cut and {\it Cut I} are collectively called as {\it Cut II} in Table~\ref{cs_ZW}. Table~\ref{cs_ZW} shows that for $m_{\hat q}=600$ GeV, $j_2$-$Z$ invariant mass cut suppress the SM background by a factor about 13, whereas, the signal is reduced by a factor of 3 only.   
\end{itemize}    

%------------------------------------------------------------------
%-------------------------------------------------------------------

\begin{figure}[t]
\begin{center}
\epsfig{file=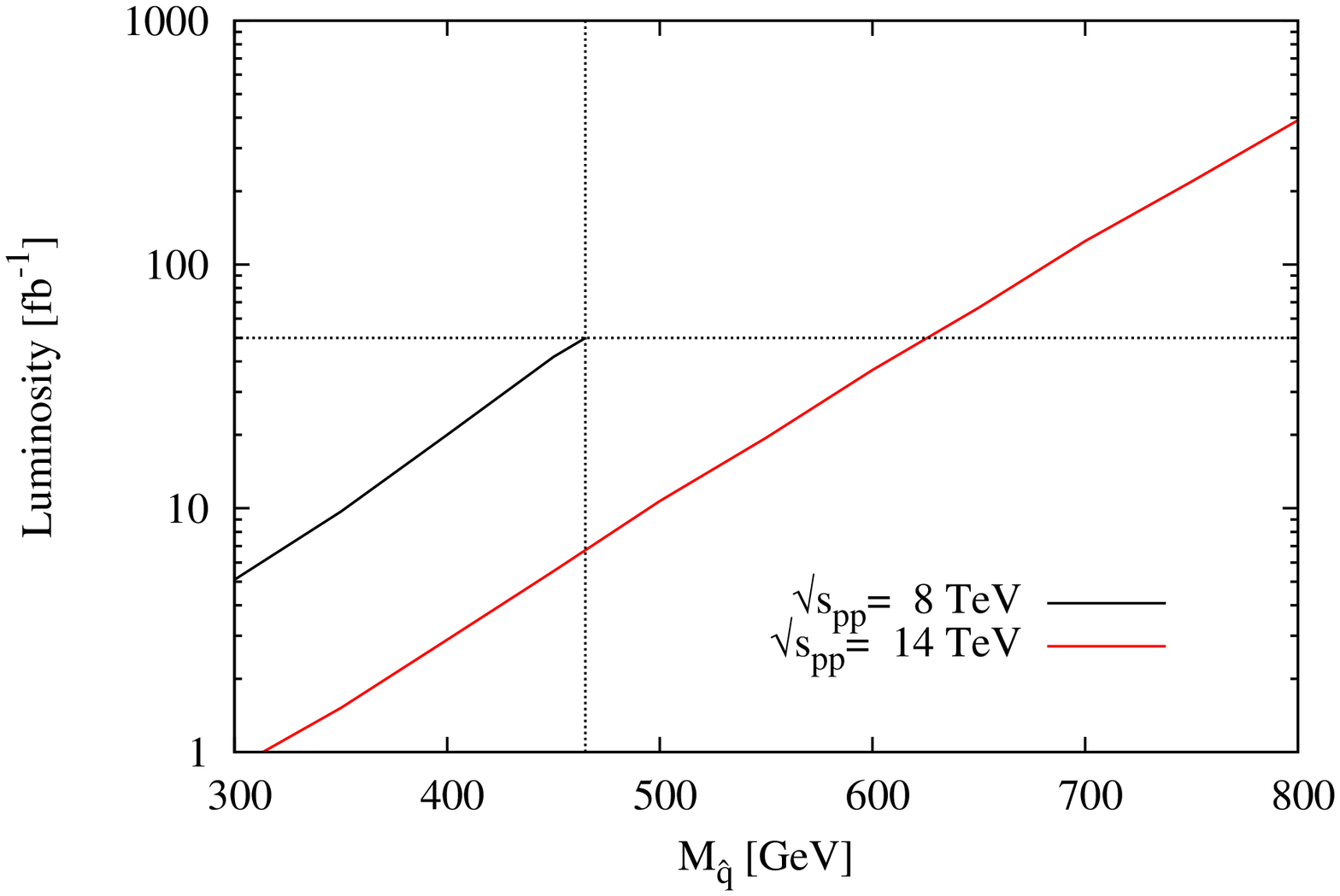,width=8cm,height=10cm,angle=-90}
\end{center}
\caption{\small{Required luminosity for $5\sigma$ discovery in {\it {two-jets}+one lepton + one $Z$-boson + $p_T\!\!\!\!\!/~$} channel is plotted as a function of $M_{\hat q}$ for the LHC with center-of-mass energy 8 TeV and 14 TeV.}}
\label{lumi_WZ}
\end{figure}

To estimate the required integrated luminosity for the discovery of the mirror quarks in {\it {two jets}+one charged lepton + $Z$ + $\slashed{p}_T$} channel, we have used Eq.~\ref{discovery}. The signal and background cross-sections after {\it Cut II} in Table~\ref{cs_ZW} shows that at the LHC with center-of-mass energy 14 TeV, 600 GeV mirror quark mass can be probed with integrated luminosity 25 fb$^{-1}$. In Fig.~\ref{lumi_WZ}, we have presented the required luminosity for $5\sigma$ discovery in {\it {two jets}+one charged lepton + $Z$ + $\slashed{p}_T$} channel as a function of $M_{\hat q}$ for the LHC with center-of-mass energy 8 TeV and 14 TeV.

%------------------------------------------------------------------
%-------------------------------------------------------------------

\section{Summary and Conclusions}\label{sec:summary}

In this work, we have a realistic left-right symmetric model with mirror fermions and mirror Higgs, and the possibility of discovering the low lying mirror fermions at the LHC. The model is $SU(3)_C \otimes SU(2)_L\otimes SU(2)_R \otimes U(1)_Y^\prime$ supplemented by a discrete $Z_2$. For each chiral multiplet of the SM fermions, we have corresponding mirror fermions of opposite chirality. The symmetry is broken to the usual SM symmetry by a mirror Higgs doublet.  The mixing between the SM fermions and the mirror fermions is achieved by using a Higgs multiplet which is a singlet under the gauge symmetry, but odd under the $Z_2$ symmetry. The model has singlet right handed neutrinos, and the corresponding mirror neutrinos which are even under $Z_2$. These are used to generate tiny neutrino masses $\simeq 10^{-11}$ GeV with a primary symmetry breaking scale of $\simeq 10^7$ GeV (which is the VEV of the mirror Higgs doublet). In this model, only the mirror fermion of the 1st family ($\hat{e}, \hat{u}, \hat{d}$) are light with well-defined relative spectrum. All the other mirror fermions are much heavier, and well above the LHC reach. Since the model is completely left-right symmetric in the fermion sector, it is naturally anomaly free. Parity conservation, and the nature of the fermion mass matrices also provides a solution for the strong CP in the model.

The light mirror fermions, $ \hat{u}, \hat{d}$, with masses around few hundred GeV to about a TeV, can be pair produced at the LHC via their QCD color interactions. They dominantly decay to a $Z$ boson plus the corresponding ordinary fermion ($\hat{u} \rightarrow{u + Z}, \hat{d}\rightarrow {d + Z}$), or to a W boson and the corresponding ordinary fermions ($\hat{u} \rightarrow{d + W}, \hat{d}\rightarrow {u + W}$). (The decays ($\hat{u} \rightarrow{u + H}, \hat{d}\rightarrow {d + H}$) are highly suppressed for most of the parameter space). Thus the most striking signal of the model is the existence of 
of resonances in the jet plus $Z$ channel. Since both the jet and the $Z$ is coming from the decay of a very heavy particle, both will have very high $p_T$. We have shown that putting a high $p_T$ cut on the jet, and reconstructing the $Z$ in the $e^+ e^-$ or $\mu^+ \mu^-$ channels, these resonances 
$\hat{u}, \hat{d}$ can be reconstructed upto  a mass of $\simeq 350$ GeV at the 8 TeV LHC, and upto  a mass of $\simeq 550$ GeV at the 14 TeV LHC. We are not aware of any other model which predicts such a resonance. We have also studied, in some detail, the final states arising from the pair productions of these light mirror fermions at the LHC. These final states are $(u Z) (\bar{u} Z), (d Z) (\bar{d} Z), (u Z) (\bar{d} W), (d Z) (\bar{u} W)$, and the subsequent decays of $W$ and $Z$ into the leptonic channels. The signals are much more observable in the (jet jet $Z Z$) channel than the (jet jet $Z W$) channel because of the missing neutrino in the latter. (The {resonance in the} signals involving the two W's will be difficult to observe). We have studied these final states and the corresponding backgrounds, and find that the reaches for the light mirror quarks can be$\simeq 450$ GeV at the 8 TeV LHC with  luminosity of $ ~30~ fb^{-1}$, and upto $ ~ 750 $ GeV  at the $14$ TeV LHC with $~300 ~fb^{-1}$ luminosity.

Our model predicts a definite pattern of spectrum for the light mirror fermions, $\hat{e}, \hat{u}, \hat{d}$. Thus with $m_{\hat{u}}< m_{\hat{d}}$, if a resonance $\hat{u}$ is observed, we expect a nearby $\hat{d}$ within few hundred GeV. This makes the prediction of the model somewhat unique. Also the $\hat{e}$ will have even lower mass, and can be looked for in the proposed future $e^+ e^{-}$ collider.

\section{Acknowledgement}

The work of SC, KG and SN is supported by US Department of Energy, Grant Number DE-FG02-04ER41306. The work of SKR was partially supported by funding available from the Department of Atomic Energy, 
Government of India, for the Regional Centre for Accelerator-based Particle Physics, Harish-Chandra 
Research Institute.

\end{document}